\documentclass[a4paper,11pt]{article}
\pdfoutput=1
\usepackage{jheppub}

\usepackage{amsmath}
\usepackage{mathtools}
\usepackage{amssymb}
\usepackage{amsthm}
\usepackage{physics}
\usepackage{braket}
\usepackage{tikz-cd}
\usepackage{dsfont}
\usepackage{graphicx}
\usepackage{subcaption}

\usepackage[normalem]{ulem}

\usepackage{orcidlink}

\numberwithin{equation}{section}

\renewcommand{\thefigure}{\arabic{section}.\arabic{figure}}

\def\be{\begin{equation}}
\def\ee{\end{equation}}

\newcommand{\bear}{\begin{eqnarray}}
\newcommand{\bea}{\begin{eqnarray}}
\newcommand{\eear}{\end{eqnarray}}
\newcommand{\eea}{\end{eqnarray}}

\def\bsq{\begin{subequations}}
\def\esq{\end{subequations}}

\def\II{\relax{\rm I\kern-.18em I}}

\newcommand*\diff{\mathop{}\!\mathrm{d}}

\title{\LARGE 

Near-extremal quantum cross-section for charged fields and superradiance
}

\author[a]{Panos Betzios\orcidlink{0000-0002-5350-9404},}
\affiliation[a]{\href{https://www.ugent.be/we/physics-astronomy/en}{Department of Physics and Astronomy},
Ghent University, \\ Krijgslaan, 281-S9, 9000 Gent, Belgium}

\author[b]{Olga Papadoulaki\orcidlink{0000-0001-5302-2930},}

\author[c]{Yanjun Zhou\orcidlink{0009-0005-9463-4155}}
\affiliation[b,c]{\href{https://www.cpht.polytechnique.fr/?q=en}{CPHT, CNRS, École polytechnique, Institut Polytechnique de Paris},  91120 Palaiseau, France}
 
\emailAdd{panos.betzios@ugent.be}
\emailAdd{olga.papadoulaki@polytechnique.edu}
\emailAdd{yanjunzhou2001@gmail.com}

\abstract{We study the scattering and absorption properties of charged scalar fields on a near-extremal Reissner-Nordstr\"om black hole background. We show that in this low-temperature regime the near-horizon throat experiences large quantum fluctuations, whose leading contribution is described by the one dimensional Schwarzian effective action, while the soft $U(1)$ gauge modes can only contribute to subleading order. We investigate the role of the leading quantum effects both inside and outside the superradiant regime. These effects result in an enhanced reflection coefficient within the superradiant regime, while causing a suppression in the non-superradiant regime. On the other hand, the absorption cross-section increases in both regimes. Additional physical features appear as kinks in the reflection coefficient and absorption cross-section plots, corresponding to the shutdown of absorption in the superradiant regime and of stimulated emission in the non-superradiant regime.}


\makeatletter
\def\@fpheader{\vspace{0cm}}
\makeatother

\begin{document}
\maketitle

\section{Introduction}
\label{sec:Introduction}

Black holes (BHs) behave as thermodynamic systems, characterized by a temperature set by their surface gravity and an entropy proportional, at leading order, to the horizon area measured in Planck units. Quantum fluctuations of gravitons and matter fields around the BH background induce corrections to this entropy \cite{Sen:2012kpz, Sen:2012cj, Sen:2012dw}. These quantum effects are particularly significant for near-extremal BHs. We now understand that such low temperature BHs do not exhibit an actual large degeneracy of states, but rather behave as ordinary quantum systems \cite{Ghosh:2019rcj, Iliesiu:2020qvm, Iliesiu:2022onk}. The key insight which led to this understanding was to realize the presence of the so-called extremal zero modes located in the near-horizon region of near-extremal BHs. The fluctuation determinant of these modes suppresses the saddle point contribution in the gravitational path integral, thereby bringing down the degeneracy of states one finds from a purely classical analysis. This understanding also allows one to study quantum corrections to Hawking radiation\footnote{Earlier works \cite{Mertens:2019bvy, Blommaert:2020yeo} have studied the quantum gravitational effects on the thermal Unruh population of BHs in JT gravity, including Euclidean wormhole contributions. These works are the 2D analogues of more recent studies on 4D BHs.} \cite{Brown:2024ajk, Lin:2025wof, Maulik:2025hax, Mohan:2024rtn, Bai:2023hpd}, absorption cross-section \cite{Emparan:2025sao, Biggs:2025nzs, Emparan:2025qqf}, decoherence of quantum superpositions \cite{Li:2025vcm}, and quasi-normal modes \cite{Jiang:2025cyl} of such BHs. It turns out that in quite a few cases, quantum corrections can lead to drastic modifications of the naive semiclassical results at sufficiently low temperatures.

In this work, we investigate the quantum corrections to the scattering and absorption cross-section of near-extremal Reissner–Nordström (RN) BHs. Unlike related studies \cite{Emparan:2025sao, Biggs:2025nzs, Emparan:2025qqf}, which focus exclusively on neutral particle absorption, our analysis specifically addresses the absorption and scattering of charged particles, which allows us to describe interesting effects such as a quantum-corrected version of superradiance \cite{Brito:2015oca}.

In fact there is an inherent connection between the Schwinger effect (recently studied in~\cite{Brown:2024ajk}) and charged superradiance. This follows from the principle of detailed balance, that relates spontaneous and stimulated emission processes of quantum mechanical systems as encapsulated in the so-called Einstein coefficients and relations between them. In the present context, the Schwinger effect describes the spontaneous charged radiation processes due to the strong electric field at the horizon, leading to a discharge of the RN BH~\cite{Gibbons:1975kk,Brown:2024ajk,Lin:2024jug}. On the other hand, superradiance is related to the ``stimulated'' scattering processes that we study in this work and has a classical manifestation in terms of wave amplification~\cite{Brito:2015oca}. Both effects are microscopically governed by the same length scale ---the Compton wavelength of the charged particle outside the BH horizon.

To analyze the absorption of charged particles, it is necessary to work within the grand canonical ensemble, so that the BH is allowed to absorb and emit charge. Compared to its RN-AdS cousin, the RN-flat\footnote{In the rest we suppress the adjective flat for simplicity.} grand canonical ensemble remains relatively underexplored\footnote{A partial reason for this is that in contrast to the AdS case, where the background behaves as a ``box'', thermodynamics in flat space are much more subtle to define and control, see \cite{Braden:1990hw, Whiting:1988qr}
for some efforts using a flat space cutoff.}. Therefore, in this work, we present a detailed study of the near-extremal RN grand canonical ensemble, laying the groundwork for a comprehensive investigation of charged particle absorption.

The remainder of this work is organized as follows. In section~\ref{sec:The Statistical Mechanics of Near-Extremal RN BHs}, we review the properties of near-extremal RN BHs and their thermodynamics, focusing on the flat space case. In section~\ref{subsec:The Canonical Ensemble} we review the quantum corrections to the near-extremal RN canonical partition function, focusing on the Schwarzian modes that generate these corrections. In section~\ref{subsec:The Grand Canonical Ensemble}, we analyze the quantum corrections to the RN grand canonical partition function. Interestingly, certain extremal zero modes present in the RN-AdS background seem to be absent in the RN case, and we offer a clarification and resolution to this puzzle. Section~\ref{sec:The RN BH Interacting with a Charged Scalar} contains most of our new results. In particular we analyze the interaction of the near-extremal RN BH with a charged scalar field and present the computation of the quantum-corrected scattering and absorption cross-section. This analysis contains three parts: In section~\ref{subsec:Computing the Ratio abs{c_{omega00}/a_{omega00}}^2} we study the fluctuation equation of the charged scalar (that takes a confluent Heun form) using the semi-classical limit of the Belavin-Polyakov-Zamolodchikov (BPZ) equation in 2D CFTs. This allows one to analytically determine the greybody factor for the charged scalar on the RN background. In section~\ref{subsec:Determining the Normalization N^2}, we combine this result with the near-horizon quantum dynamics that involves the computation of the near-horizon quantum correlators for the charged scalar (its harmonics on the transverse $S^2$ lead to a collection of CFT operators whose correlators can be computed via the effective 1D Schwarzian theory, as we review in appendix~\ref{subsec:The Density of States and Matter Correlators}). An overall normalization constant can be fixed by matching the combined result to the known semi-classical result for Hawking radiaton. In section~\ref{subsec:Plots of abs{R}^2 and sigma_{abs}} we present our results for the quantum corrected reflection coefficient and absorption cross-section both inside and outside the superradiant regime for the s-wave sector. Additionally, in section~\ref{subsec:Comments on Higher Partial Waves} we carry the same analysis for higher partial waves and comment on their physical difference with respect to the dominant s-wave case. Finally, in the discussion section~\ref{sec:Discussions}, we summarize our findings and suggest several directions for future work.

\section{The statistical mechanics of near-extremal RN BHs}
\label{sec:The Statistical Mechanics of Near-Extremal RN BHs}

The Euclidean RN BH is a classical solution to the Euclidean Einstein-Maxwell theory
\begin{equation}
    I_\text{EM}[g,A]=\int_{ \mathcal{M}} \diff^4x\sqrt{g}\,\left(-\frac{1}{16\pi G_N}R+\frac{1}{4e^2}F^2\right) - \frac{1}{8\pi G_N} \int_{\partial \mathcal{M}} \diff^3 x \sqrt{\gamma} \,K_{GH} \,,
\end{equation}
where $R$ is the Ricci scalar, $F=\diff A$ is the $U(1)$ gauge field strength, $G_N$ is the Newton’s constant, and $e$ is the gauge field coupling constant. The last term is the Gibbons-Hawking boundary term, needed for a well defined variational problem.

An electrically charged RN solution is parameterized by its mass $M$ and electric charge $Q$. The metric (in Euclidean signature) is given by
\begin{equation}
   \diff s^2_\text{RN} = g^\text{RN}_{\mu \nu} \diff x^\mu \diff x^\nu = f(r)\diff\tau^2+\frac{\diff r^2}{f(r)}+r^2\diff\Omega_2^2,\quad f(r)=1-\frac{2G_N M}{r}+\frac{G_N}{4\pi} \frac{Q^2}{r^2}\,,
\end{equation}
from which we can read off the horizon radius $r_+$ as the larger root of $f(r)=0$. The gauge field is given by
\begin{equation}
    A_\text{RN}=i\mu\left(1-\frac{r_+}{r}\right)\diff \tau,\quad \mu=\frac{e}{4\pi}\frac{Q}{r_+}\,,
\end{equation}
where $\mu$ is the chemical potential defined as the potential difference between the infinity and the horizon. The inverse Hawking temperature $\beta=T^{-1}$ of the BH is
\begin{equation}
    \beta=\frac{4\pi}{f'(r_+)}\,.
\end{equation}
The extremal BH has zero temperature. In this work, we focus on the near-extremal regime in which the temperature is low. We will shortly see that in this regime, semiclassical thermodynamics break down and quantum effects are important.

Fluctuations of $g_{\mu \nu}$ and $A$ around the RN BH background 
\begin{equation}
    g_{\mu \nu} =  g^\text{RN}_{\mu \nu} + h_{\mu \nu}, \quad A = A_\text{RN}+\mathcal{A}\,,
\end{equation}
result in quantum corrections to the RN partition function. By expanding the Einstein-Maxwell action to quadratic order in the fluctuations $h_{\mu \nu}$ and $\mathcal{A}$, we get
\begin{equation}
    I_\text{EM}[g,A]=I_\text{EM}[g_\text{RN},A_\text{RN}]+\int\diff^4x\sqrt{g_\text{RN}}\,\Psi\Delta\Psi+\cdots\,.
\end{equation}
The zeroth order term is the on-shell action for the RN BH background. After including boundary counterterms to render it finite\footnote{There is always a ``scheme'' ambiguity in the form of finite contributions. What is invariant and scheme independent are differences of the on-shell action between backgrounds i.e. $I_\text{EM}[g_\text{RN},A_\text{RN}] - I_\text{EM}[g_\text{flat},A_\text{flat}]$.}, it is the grand potential of the RN BH
\begin{equation}
    \Omega_\text{RN}(\beta,\mu)=\frac{I_\text{EM}[g_\text{RN},A_\text{RN}]}{\beta}\,.
\end{equation}
The first-order term vanishes as the background satisfies the equations of motion. For our purposes, the precise expression \cite{Bhattacharyya:2012wz} for the second-order term is not important, thus we let $\Psi$ represent fluctuations, and $\Delta$ is a 2-derivative differential operator constructed out of the background. By integrating out the quadratic action of fluctuations, we obtain the one-loop corrected RN grand canonical partition function
\begin{equation}
    \mathcal{Z}_\text{RN}[\beta,\mu]=\int\mathcal{D}g\mathcal{D}A\,e^{-I_\text{EM}[g,A]}\approx \underbrace{e^{-\beta\Omega_\text{RN}(\beta,\mu)}}_{\text{zero-loop}}\underbrace{(\det\Delta)^{-\frac{1}{2}}}_{\text{one-loop}}\,.
\end{equation}

In the near-extremal regime, we can expand the operator $\Delta$ around its extremal counterpart $\Delta^{(0)}$
\begin{equation}
    \Delta=\Delta^{(0)}+T\Delta^{(1)}+O(T^2)
\end{equation}
Eigenvalues $\{\Lambda_n\}$ of $\Delta$ admit a similar expansion $\Lambda_n=\Lambda_n^{(0)}+T\Lambda_n^{(1)}+O(T^2)$, where $\{\Lambda_n^{(0)}\}$ are the eigenvalues of the extremal operator $\Delta^{(0)}$. Following \cite{Banerjee:2023quv}, we call the eigenfunctions of $\Delta$ with $\Lambda_n^{(0)}=0$ but $\Lambda_n^{(1)}\neq 0$ extremal zero modes, those with $\Lambda_n^{(0)}\neq 0$ extremal non-zero modes, and those with $\Lambda_n^{(0)}=\Lambda_n^{(1)}= 0$ near-extremal zero modes. We cannot perform the Gaussian integral over the near-extremal zero modes when the expansion in $T$ is truncated at $O(T^2)$. Nevertheless, in \cite{Banerjee:2023quv}, it is shown that the measure for these modes contributes at $O(1)$ to $\mathcal{Z}_\text{RN}$ rather than at $O(T)$. Therefore, only extremal zero modes contribute important $\log T$ terms to $\log \mathcal{Z}_\text{RN}$. Contributions from extremal non-zero modes and near-extremal zero modes are both of the form $\log\#+O(T)$, where $\#\sim O(1)$, and so they are polynomially suppressed in $T$. We should emphasize that $\log T$ quantum corrections are of utmost importance for near-extremal BHs, since they compete with the classical terms at low enough temperatures.

A direct evaluation of the near-extremal one-loop partition function by solving for extremal zero modes is involved. In \cite{Iliesiu:2022onk, Banerjee:2023gll}, the $\log T$ corrections to the logarithm of the RN canonical partition function are computed analytically. In \cite{Kolanowski:2024zrq}, the corrections to the BTZ grand canonical partition function are computed analytically, while the corrections to the RN-AdS grand canonical partition function are studied numerically. However, an attempt in computing them for the RN (flat) grand canonical partition function is missing. The reason is that the near-extremal RN grand canonical ensemble suffers from a few pathological issues, as we will shortly describe. Nevertheless, \cite{Kolanowski:2024zrq} proposed what the extremal zero modes for the RN grand canonical ensemble should be, based on their BTZ and RN-AdS counterparts. In this work, we will elaborate on their proposal and apply it to study quantum corrections to the scattering and absorption cross-section of a charged scalar field on the near-extremal RN BH background.

\begin{figure}
    \centering
    \includegraphics[width=0.6\linewidth]{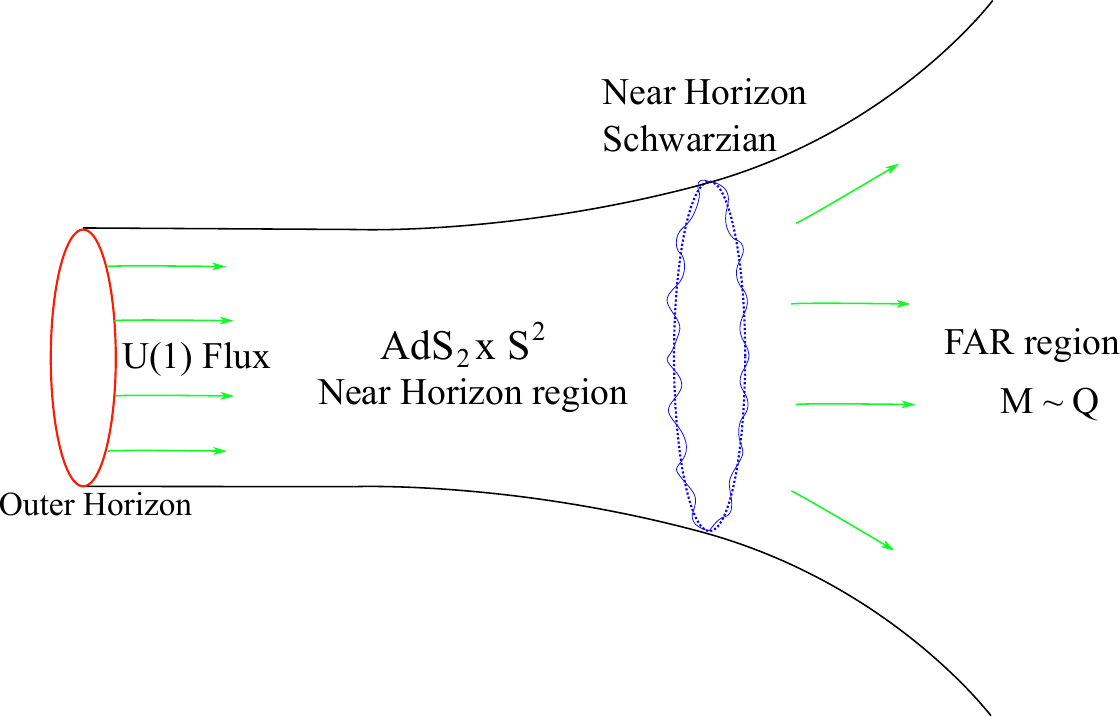}
    \caption{The exterior of a near-extremal RN BH. The near horizon region is approximately $AdS_2 \times S^2$ supported by electric flux. From far away the RN BH has mass approximately equal to its charge.}
    \label{fig:throat}
\end{figure}

There is a 1D effective theory that captures the one-loop corrections to the partition function of higher-dimensional near-extremal BHs. To be concrete, we focus here on the RN BH. The near-horizon geometry of a near-extremal RN BH is approximately AdS$_2\times S^2$. This near-horizon region is called the throat, see Figure~\ref{fig:throat}. Quantum fluctuations are large inside the throat, but small outside. To obtain the 1D effective theory, we first do a Kaluza-Klein reduction on $S^2$ to obtain a 2D effective theory. Then we separate the integral in the 2D effective action into the inside-throat part and the outside-throat part. The outside-throat part can be approximated by the on-shell action evaluated at the near-extremal solution. The inside-throat part can be approximated by an effective JT gravity action. The entire dynamics of JT gravity are contained in soft boundary modes captured in its boundary term, resulting in the 1D effective theory that allows us to compute the quantum-corrected BH density of states and the transition rates between different BH states\footnote{From the above discussion, it is evident that extremal zero modes play the central role in near-extremal physics. A specific set of such modes (the Schwarzian modes) also appear in JT gravity as its soft boundary modes. As noted in \cite{Banerjee:2023quv}, however, the precise equivalence between Einstein–Maxwell theory around a near-extremal BH and JT gravity remains questionable. Nevertheless, we will adopt the dimensional-reduction route described above as a convenient way to obtain the effective 1D Schwarzian theory.}.

This procedure is explained in detail in the literature. For the RN-AdS grand canonical ensemble and the RN canonical ensemble, see \cite{Iliesiu:2020qvm}\footnote{Complementary approaches to deriving the 1D effective action for the RN-AdS grand canonical ensemble have been developed in \cite{Sachdev:2019bjn, Moitra:2018jqs}. In particular, \cite{Moitra:2018jqs} showed that this effective action also captures the response of a near-extremal RN-AdS BH to a probe charged scalar at low energies. This work builds on \cite{Nayak:2018qej}, where the corresponding 1D effective action capturing the response to a probe neutral scalar was derived.}. For the BTZ grand canonical ensemble, see \cite{Ghosh:2019rcj}. We are interested in the RN grand canonical ensemble---in which the BH charge is not fixed---because we are going to study the absorption cross-section of the RN BH for a charged scalar field, thus there will be exchange of charge between them. In the next two subsections, we will first briefly review the RN canonical ensemble, and then propose the correct 1D effective action governing the soft modes for the RN grand canonical ensemble that contribute to leading order.

\subsection{The canonical ensemble}
\label{subsec:The Canonical Ensemble}

In the RN canonical ensemble, we fix the temperature $T=\beta^{-1}$ and the BH charge $Q$. Without loss of generality, we assume $Q>0$. The extremal horizon radius and energy are respectively
\begin{equation}
\label{eqn:RN extremal quantities}
    r_0(Q)=r_+(T=0,Q)=\sqrt{\frac{G_N}{4\pi}}Q,\quad M_0(Q)=M(T=0,Q)=\frac{Q}{\sqrt{4\pi G_N}}\,.
\end{equation}

In the near-horizon region (NHR) defined by $\rho= r-r_0\ll r_0$, the near-extremal RN BH develops an AdS$_2\times S^2$ throat
\begin{equation}
    g_\text{RN}\approx \underbrace{\frac{\rho^2-\delta r_h^2}{r_0^2}\diff\tau^2+\frac{r_0^2}{\rho^2-\delta r_h^2}\diff\rho^2}_{\text{the BH patch of AdS$_2$ with AdS radius $r_0$}}+\underbrace{(r_0+\rho)^2\diff\Omega_2^2}
    _{\substack{\text{an $S^2$ with radius}\\\text{approximately $r_0$}}}
\end{equation}
where we have introduced $\delta r_h= r_+-r_0$. The far-away region (FAR) is located at large $r$, where the geometry is approximately extremal
\begin{equation}
    g_\text{RN}\approx f_0(r)\diff\tau^2+\frac{\diff r^2}{f_0(r)}+r_0^2\diff\Omega_2^2,\quad f_0(r)=\frac{(r-r_0)^2}{r^2}\,.
\end{equation}
In the NHR, quantization is easy and necessary. In the FAR, quantization is hard, but quantum fluctuations are suppressed. These two regions overlap in the bulk. We will deal with them separately and then perform an appropriate matching at the boundary of the NHR ($\partial\text{NHR}$).

We first perform a dimensional reduction from 4D to 2D by integrating out the $S^2$ degrees of freedom. We consider the ansatz for the 4D metric to be
\begin{equation}
    g_4=\frac{r_0}{\chi^\frac{1}{2}}(g_2)_{\mu\nu}\diff x^\mu\diff x^\nu +\chi \diff\Omega_2^2\,,
\end{equation}
where $x^\mu=(\tau,\rho)$ are AdS$_2$ coordinates, and $\chi(x)$ is the dilaton that parameterizes the size of the $S^2$. After the reduction, the Einstein-Maxwell theory becomes a 2D effective theory of dilaton gravity
\begin{equation}
\label{eqn:dilation gravity action}
    I_Q[g_2,\chi]=-\frac{1}{4G_N}\int \diff^2x\sqrt{g_2}\,\left(\chi R_2-2U_Q(\chi)\right)-\frac{1}{2G_N}\int_\partial \diff u\sqrt{h_1}\,\chi K_1\,,
\end{equation}
where the dilation potential is
\begin{equation}
    U_Q(\chi)=r_0\left(\frac{G_N Q^2}{4\pi \chi^\frac{3}{2}}-\frac{1}{\chi^\frac{1}{2}}\right)\,.
\end{equation}
The solution to the equations of motion is
\begin{equation}
    \chi(r)=r^2,\quad g_2=\frac{\chi^\frac{1}{2}}{r_0}\left(f(r)\diff\tau^2+\frac{\diff r^2}{f(r)}\right),\quad f(r)=1-\frac{2G_N M}{r}+\frac{G_N}{4\pi}\frac{Q^2}{r^2} \, ,
\end{equation}
showing that the 2D part of the RN BH metric remains a solution of the reduced action, up to a conformal factor parametrised by the dilaton.

Although a direct quantization of this dilaton gravity model is currently out of reach, we can progress further by separating the integral in \eqref{eqn:dilation gravity action} into the NHR part and the FAR part. In the NHR where $\frac{r-r_0}{r_0}\ll 1$, we can study small
fluctuations of the dilaton
\begin{equation}
    \frac{\chi(r)}{G_N}=\Phi_0+\Phi(r)\,,
\end{equation}
around its extremal value
\begin{equation}
\label{eqn:RN extremal entropy}
    \Phi_0=\frac{\chi(r_0)}{G_N}=\frac{r_0(Q)^2}{G_N}\,.
\end{equation}
Expanding \eqref{eqn:dilation gravity action} to $O(\Phi/\Phi_0)$, we get
\begin{equation}
\label{eqn:NHR action}
    I_Q^\text{NHR}[g_2,\chi]\approx -\frac{1}{4}\int_\text{NHR}\diff^2x\sqrt{g_2}\,\left(\Phi_0 R_2+\Phi\left(R_2+\frac{2}{r_0^2}\right)\right)\,.
\end{equation}
In the FAR region, we expand \eqref{eqn:dilation gravity action} around the extremal solution to first order in fluctuations. Higher order fluctuations in the FAR region are suppressed compared to \eqref{eqn:NHR action}. Following from the variational principle, the first order term is a total derivative and so induces a boundary term on $\partial\text{NHR}$. See~\cite{Iliesiu:2020qvm} for details.

In total one finds,
\begin{equation}
    I_Q^\text{FAR}[g_2,\chi]\approx\beta M_0(Q)-\frac{1}{2}\int_{\partial\text{NHR}}\diff u\sqrt{h_1}\,\left(\Phi_0 K_\text{NHR}+\frac{2 r_0^3}{G_N \epsilon} \left( K_\text{NHR}-\frac{1}{r_0}\right)\right)\,,
\end{equation}
where $\epsilon$ is some parameter fixed by the proper length $\ell$ of $\partial\text{NHR}$ via $\ell=\beta r_0/\epsilon$.

The RN canonical partition function in the near-extremal regime is thus described by
\begin{equation}
    Z_\text{RN}[\beta,Q]=e^{-\beta M_0(Q)+S_0(Q)}\int\mathcal{D}g_2\mathcal{D}\Phi\, e^{-I_\text{JT}[g_2,\Phi]}\,,
\end{equation}
where we have used the Euler characteristic of the disk
\begin{equation}
    \frac{1}{4}\int_\text{NHR}\diff^2x\sqrt{g_2}\, R_2+\frac{1}{2}\int_{\partial\text{NHR}}\diff u\sqrt{h_1}\,K_\text{NHR}=\pi\,,
\end{equation}
and have identified $\pi \Phi_0$ with the extremal entropy $S_0$ via the Bekenstein–Hawking formula. The action of the JT gravity is
\begin{equation}
    I_\text{JT}[g_2,\Phi]=-\frac{1}{4}\int_\text{NHR}\diff^2x\sqrt{g_2}\,\Phi\left(R_2+\frac{2}{r_0^2}\right)-\frac{r_0^3}{G_N \epsilon}\int_{\partial\text{NHR}}\diff u\sqrt{h_1}\,\left( K_\text{NHR}-\frac{1}{r_0}\right)\,.
\end{equation}
The path integral over the dilaton $\Phi$ enforces $R_2=-2/r_0^2$. Thus, each NHR configuration describes a patch of AdS$_2$ cut along a curve with a fixed proper length $\ell$. In the Poincar\'e coordinates, the AdS$_2$ metric is
\begin{equation}
    g_2=r_0^2\frac{\diff F^2+\diff z^2}{z^2}\,.
\end{equation}
We parametrize its boundary curve $(F(u),z(u))$ with a proper time $u$, with $u\in[0,\beta)$ and $(h_1)_{uu}=r_0^2/\epsilon^2$. Then the boundary metric
\begin{equation}
    h_1=r_0^2\frac{F'(u)^2+z'(u)^2}{z^2}\diff u^2=\frac{r_0^2}{\epsilon^2}\diff u^2\,,
\end{equation}
imposes $z(u)\approx \epsilon F'(u)$ in the $\epsilon\ll \beta$ limit. The extrinsic curvature can then be written in terms of the Schwarzian derivative
\begin{equation}
    K_\text{NHR}\approx \frac{1}{r_0}\left(1+\epsilon^2\left\{F(u),u\right\}\right),\quad  \left\{F(u),u\right\}=\frac{F'''(u)}{F'(u)}-\frac{3F''(u)^2}{2F'(u)^2}\,.
\end{equation}
Now the RN canonical partition function reduces to a path integral over the field $F$ (modulo $SL(2,\mathbb{R})$, which is a gauge redundancy of the Schwarzian action)
\begin{equation}
    Z_\text{RN}[\beta,Q]=e^{-\beta M_0(Q)+S_0(Q)}\int\frac{\mathcal{D}F}{SL(2,\mathbb{R})}\,\exp(\frac{r_0^3}{G_N}\int_0^\beta\diff u\,\left\{F(u),u\right\})\,.
\end{equation}
This path integral can be computed exactly and captures the one-loop correction due to the Schwarzian modes, which are extremal zero modes that are universally present in all near-extremal BH spacetimes.

The geometry we are working with in the NHR after reducing on $S^2$ is actually the hyperbolic disk inherited from the topology of the Euclidean BH. We can go from the Poincar\'e coordinates to the disk by replacing
\begin{equation}
    F(u)=\tan\frac{\pi\tau(u)}{\beta},\quad \tau(u+\beta)=\tau(u)+\beta,\quad \tau'(u)\geq 0\,,
\end{equation}
so that $\tau(u)$ can be interpreted as a time reparametrization of $\partial\text{NHR}$. We define 
\begin{equation}
\label{eqn:RN Schwarzian coupling}
    C(Q)=\frac{r_0(Q)^3}{G_N} \, ,
\end{equation}
and compute the Schwarzian path integral
\begin{equation}
    Z_\text{Sch}(C(Q),\beta)=\int\frac{\mathcal{D}\tau}{SL(2,\mathbb{R})}\,\exp(C\int_0^\beta\diff u\,\left\{\tan\frac{\pi\tau(u)}{\beta},u\right\})\,,
\end{equation}
by expanding $\tau(u)=u+\epsilon(u)$ in terms of its saddle $\tau_\text{cl}(u)=u$ plus fluctuations $\epsilon(u)=\epsilon(u+\beta)$. It is convenient to expand the fluctuations in Fourier space
\begin{equation}
    \epsilon(u)=\frac{\beta}{2\pi}\sum_{n\neq -1,0,1}e^{-\frac{2\pi}{\beta}inu}\epsilon_n+\text{h.c.}\,,
\end{equation}
where we have removed the $n=-1,0,1$ modes because of the modding by $SL(2,\mathbb{R})$ \cite{Maldacena:2016upp}. The $\epsilon_n$'s $(n\neq-1,0,1)$ are identified with the extremal Schwarzian zero modes \cite{Iliesiu:2022onk}. The Schwarzian action to quadratic order is given by
\begin{equation}
    C\int_0^\beta\diff u\,\left\{\tan\frac{\pi\tau(u)}{\beta},u\right\}\approx\frac{2\pi^2 C}{\beta}+\frac{4\pi^2 C}{\beta}\sum_{n>1}n^2(n^2-1)\Bar{\epsilon}_n\epsilon_n\,.
\end{equation}
Using a zeta-function regularization procedure, the one-loop corrected Schwarzian path integral is
\begin{equation}
\label{eqn:Schwarzian path integral}
    Z_\text{Sch}(C(Q),\beta)=\frac{1}{4\pi^2}\left(\frac{2\pi C}{\beta}\right)^\frac{3}{2}e^{\frac{2\pi^2 C}{\beta}}\,.
\end{equation}
In fact, the Schwarzian path integral is one-loop exact \cite{Stanford:2017thb}, so \eqref{eqn:Schwarzian path integral} is the entire answer.

The one-loop corrected BH energy and entropy can be read off from the one-loop corrected RN free energy $-\beta F_\text{RN}=\log Z_\text{RN}$. 
\begin{equation}
    S(\beta,Q)=(1-\beta\partial_\beta)\log Z_\text{RN}=S_0+\frac{4\pi^2 C}{\beta}-\frac{3}{2}\log\frac{\beta}{eC}\,,
\end{equation}
\begin{equation}
\label{eqn:corrected RN canonical energy}
    M(\beta, Q)=F_\text{RN}+TS=M_0+\frac{2\pi^2 C}{\beta^2}+\frac{3}{2\beta}\,.
\end{equation}

The first terms are extremal quantities. The second terms are their leading semiclassical corrections away from extremality. The third terms are their one-loop corrections which dominate at low temperatures $C/\beta\ll 1$. Therefore, the breakdown energy scale for semiclassical thermodynamics is $E_\text{break} = 1/C$.

\subsection{The grand canonical ensemble}
\label{subsec:The Grand Canonical Ensemble}

In the RN grand canonical ensemble, we fix the temperature $T=\beta^{-1}$ and the chemical potential $\mu$ and allow fluctuations and charge exchange of the BH with its environment.

In this subsection, we first review the near-extremal one-loop corrected RN-AdS grand canonical partition function studied in \cite{Iliesiu:2020qvm}. If we take the AdS$_4$ radius $L$ to infinity, the cosmological constant $\Lambda=-3/L^2$ vanishes. Therefore, $L\to\infty$ is called the flat-space limit. One might expect that all RN-AdS results reduce to their RN counterparts in this limit. However, we will point out a subtlety in this naive expectation. Nevertheless, we will propose a modification of the RN-AdS analysis in order to study the RN grand canonical ensemble. The RN grand canonical ensemble suffers from a few pathological issues, which make it less straightforward than its RN-AdS counterpart. Some of these issues also appear in the BTZ grand canonical ensemble, but the BTZ case is much better understood. Thus, we will motivate our RN analysis by first explaining how the same issues are resolved in the BTZ case.

Following exactly the same procedure as in section~\ref{subsec:The Canonical Ensemble}, the near-extremal one-loop corrected RN-AdS canonical partition function is 
\begin{equation}
    Z_\text{RN-AdS}[\beta, Q]=e^{-\beta \Tilde{M}_0(Q)+\Tilde{S}_0(Q)}Z_\text{Sch}(\Tilde{C}(Q),\beta)\,,
\end{equation}
where quantities with a tilde are the RN-AdS counterparts of the quantities without a tilde
\begin{equation}
\label{eqn:RN-AdS extremal quantities}
    Q^2=\frac{4\pi}{G_N}\left(\Tilde{r}_0^2+\frac{3\Tilde{r}_0^4}{L^2}\right),\quad \Tilde{M}_0=\frac{\Tilde{r}_0}{G_N}\left(1+\frac{2\Tilde{r}_0^2}{L^2}\right),\quad \Tilde{S}_0=\frac{\pi\Tilde{r}_0^2}{G_N}\,.
\end{equation}
We solve for $\Tilde{r}_0(Q)$ from the first equation of \eqref{eqn:RN-AdS extremal quantities}
\begin{equation}
\label{eqn:RN-AdS r_0(Q)}
    \tilde{r}_0^2(Q)=\frac{L^2}{6}\left(-1+\sqrt{1+\frac{3G_NQ^2}{\pi L^2}}\right)\,.
\end{equation}
Then the last two equations of \eqref{eqn:RN-AdS extremal quantities} give $\Tilde{M}_0(Q)$ and $\Tilde{S}_0(Q)$ respectively. The coupling of the Schwarzian action is
\begin{equation}
\label{eqn:RN-AdS Schwarzian coupling}
    \Tilde{C}=\frac{L^2\Tilde{r}_0^3}{G_N(L^2+6\Tilde{r}_0^2)}\,.
\end{equation}
We see that \eqref{eqn:RN-AdS extremal quantities} and \eqref{eqn:RN-AdS Schwarzian coupling} reduce to \eqref{eqn:RN extremal quantities}, \eqref{eqn:RN extremal entropy}, and \eqref{eqn:RN Schwarzian coupling} in the flat-space limit.

The near-extremal one-loop corrected RN-AdS grand canonical partition function is thus (see also~\cite{Iliesiu:2020qvm})
\begin{equation}
\label{eqn:RN-AdS grand partition function from partition function}
    \mathcal{Z}_\text{RN-AdS}[\beta,\mu]=\sum_{Q\in e\cdot\mathbb{Z}}e^{\beta\mu\frac{Q}{e}} Z_\text{RN-AdS}[\beta,Q]=\sum_{Q\in e\cdot\mathbb{Z}}e^{\beta\mu\frac{Q}{e}-\beta \Tilde{M}_0(Q)+\Tilde{S}_0(Q)}Z_\text{Sch}(\Tilde{C}(Q),\beta)\,.
\end{equation}
In the near-extremal regime $\beta\gg \Tilde{r}_0$, the sum over $Q$ is approximately dominated by the saddle point $\Tilde{Q}_0$ of the $O(\beta)$ term $\beta\mu Q/e-\beta \Tilde{M}_0(Q)$ \cite{Iliesiu:2020qvm}. Using \eqref{eqn:RN-AdS extremal quantities} and \eqref{eqn:RN-AdS r_0(Q)}, the saddle point is evaluated to be
\begin{equation}
\label{eqn:RN-AdS O(beta) saddle}
    \left.\pdv{}{Q}\left(\Tilde{M}_0(Q)-\mu\frac{Q}{e}\right)\right\rvert_{Q=\Tilde{Q}_0}=0\quad\Rightarrow\quad \Tilde{Q}_0^2=\left(\frac{4\pi L\mu}{e^2}\right)^2\frac{4\pi G_N\mu^2-e^2}{3}\,.
\end{equation}
Note that $\Tilde{Q}_0$ is also the extremal mean value of the charge in the RN-AdS grand canonical ensemble \cite{Sachdev:2019bjn}
\begin{equation}
\label{eqn:mean value of the charge in the RN-AdS grand canonical ensemble}
    \Tilde{Q}_0=-e\left.\pdv{}{\mu}\Omega_\text{RN-AdS}(\beta,\mu)\right\rvert_{T=0}\,.
\end{equation}
Although $\Tilde{Q}_0$ is not the true saddle point of the complete \eqref{eqn:RN-AdS grand partition function from partition function}, it is still useful to expand around it: $Q=\Tilde{Q}_0+e\cdot q$, such that $q\in\mathbb{Z}$. To leading order in $q$, 
\begin{equation}
    \Tilde{M}_0(Q)-\mu\frac{Q}{e}\approx \Tilde{M}_0(\Tilde{Q}_0)-\mu\frac{\Tilde{Q}_0}{e}+\frac{q^2}{2\Tilde{K}},\quad \Tilde{K}=\frac{4\pi (L^2+6\Tilde{r}_0^2)}{3e^2\Tilde{r}_0}\,.
\end{equation}
\begin{equation}
    \Tilde{S}_0(Q)\approx \Tilde{S}_0(\Tilde{Q}_0)+2\pi \Tilde{\mathcal{E}} q,\quad \Tilde{\mathcal{E}}=\frac{eL\Tilde{r}_0}{L^2+6\Tilde{r}_0^2}\sqrt{\frac{L^2+3\Tilde{r}_0^2}{4\pi G_N}}\,.
\end{equation}
In particular, the extremal energy of a charged-$Q$ RN-AdS BH is
\begin{equation}
\label{eqn:corrected RN-AdS grand canonical energy due to n}
    \Tilde{M}_0(Q)\approx \Tilde{M}_0(\Tilde{Q}_0)+\mu q+\frac{q^2}{2\Tilde{K}}\,.
\end{equation}
Just as the leading semiclassical correction $2\pi^2 CT^2$ to the extremal energy $M_0$ in \eqref{eqn:corrected RN canonical energy} arises from the deviation of the temperature from extremality, the $q^2/2\Tilde{K}$ term in \eqref{eqn:corrected RN-AdS grand canonical energy due to n} plays a similar role due to the deviation of the charge from $\Tilde{Q}_0$. 

Following \cite{Iliesiu:2020qvm}, we will choose $\mu$ such that $\Tilde{Q}_0\gg 1$. In the large $\Tilde{Q}_0$ limit,
\begin{equation}
    Z_\text{Sch}(\Tilde{C}(Q),\beta)\approx Z_\text{Sch}(\Tilde{C}(\Tilde{Q}_0),\beta)\,,
\end{equation}
and so
\begin{equation}
\label{eqn:near-extremal one-loop corrected RN grand canonical partition function}
    \mathcal{Z}_\text{RN-AdS}[\beta,\mu]\approx e^{\beta\mu\frac{\Tilde{Q}_0}{e}-\beta \Tilde{M}_0(\Tilde{Q}_0)+\Tilde{S}_0(\Tilde{Q}_0)}Z_\text{Sch}(\Tilde{C}(\Tilde{Q}_0),\beta)\sum_{q\in\mathbb{Z}}e^{2\pi\Tilde{\mathcal{E}}q-\beta\frac{q^2}{2\Tilde{K}}}\,.\end{equation}
Using the modular transformation property of the Jacobi theta function, we have the identity
\begin{equation}
\label{eqn:one-loop correction from U(1) mode}
    Z_{U(1)}(\Tilde{K},\Tilde{\mathcal{E}},\beta)=\sum_{q\in\mathbb{Z}}e^{2\pi\Tilde{\mathcal{E}}q-\beta\frac{q^2}{2\Tilde{K}}}=\sqrt{\frac{2\pi\Tilde{K}}{\beta}}e^{\frac{\Tilde{K}}{2\beta}(2\pi\Tilde{\mathcal{E}})^2}\sum_{m\in\mathbb{Z}}e^{-\frac{2\pi^2\Tilde{K}}{\beta}m^2}e^{-i\frac{2\pi^2\Tilde{K}}{\beta}\frac{2\pi\Tilde{\mathcal{E}}}{\pi}m}\,.
\end{equation}
From \eqref{eqn:one-loop correction from U(1) mode}, we can identify the energy scale $1/\Tilde{K}$ such that for $\Tilde{K}\ll\beta$, $Z_{U(1)}\approx1$\footnote{From \eqref{eqn:corrected RN-AdS grand canonical energy due to n}, we see that there is an energy gap $\mu q+\frac{q^2}{2\tilde{K}}$ between the ground state of a charged-$\tilde{Q}_0$ BH and that of a charged-$Q$ BH. In the $\tilde{K}\ll \beta$ regime, the thermal energy is insufficient to overcome this gap, effectively freezing charge fluctuations. Consequently, the system behaves as if it were in the fixed-charge canonical ensemble, and we have $Z_{U(1)}\approx1$.}, while for $\Tilde{K}\gg\beta$, $Z_{U(1)}$ contributes to the $\log T$ corrections to $\log\mathcal{Z}_\text{RN}$:
\begin{equation}
    Z_{U(1)}(\Tilde{K},\Tilde{\mathcal{E}},\beta)\approx\sqrt{\frac{2\pi\Tilde{K}}{\beta}}e^{\frac{\Tilde{K}}{2\beta}(2\pi\Tilde{\mathcal{E}})^2}\,.
\end{equation}
$Z_{U(1)}$ captures the one-loop correction due to the extremal zero modes coming from the $U(1)$ gauge field.

As shown in \cite{Mertens:2019tcm}, the complete one-loop correction can be reproduced by a 1D effective theory
\begin{equation}
\label{eqn:1D effective theory for one-loop RN-AdS}
\begin{split}
    &\mathcal{Z}^\text{one-loop}_\text{RN-AdS}[\beta,\mu]=Z_\text{Sch}(\Tilde{C}(\Tilde{Q}_0),\beta)Z_{U(1)}(\Tilde{K},\Tilde{\mathcal{E}},\beta)\\
    =&\int\frac{\mathcal{D}\tau}{SL(2,\mathbb{R})}\frac{\mathcal{D}\theta}{U(1)}\,\exp(\Tilde{C}(\Tilde{Q}_0)\int_0^\beta\diff u\,\left\{\tan\frac{\pi\tau(u)}{\beta},u\right\}-\frac{\Tilde{K}}{2}\int_0^\beta\diff u\,\left(\theta'(u)+i\frac{2\pi\Tilde{\mathcal{E}}}{\beta}\tau'(u)\right)^2)\,.
\end{split}
\end{equation}
This also matches the result of \cite{Sachdev:2019bjn, Moitra:2018jqs}. Besides the Schwarzian action, one also finds an action with a coupling $\Tilde{K}$ for the extremal $U(1)$ gauge zero modes\footnote{The fact that the one-loop partition function factorises into a Schwarzian part and a $U(1)$ part, is not evident from the path integral . The transformation that achieves this decoupling is discussed in~\cite{Mertens:2019tcm, Chaturvedi:2018uov}.}.

One might expect to obtain the near-extremal one-loop corrected RN grand canonical partition function by taking the flat-space limit of \eqref{eqn:near-extremal one-loop corrected RN grand canonical partition function}. Here we point out a subtlety involved in the process. The mean value of the horizon radius in the RN grand canonical ensemble is
\begin{equation}
\label{eqn:RN grand canonical horizon radius}
    r_+(T,\mu)=\frac{e^2-4 \pi  G_N \mu ^2}{4 \pi  e^2 T}\,,
\end{equation}
while in the RN-AdS grand canonical ensemble, it is determined by the quadratic equation
\begin{equation}
    T=\frac{1}{4\pi e^2 \Tilde{r}_+}\left(e^2\left(1+\frac{3 \Tilde{r}_+^2}{L^2}\right)-4\pi G_N \mu ^2\right)\,.
\end{equation}
The solution in the RN-AdS case has the following two branches
\begin{equation}
    \Tilde{r}_+(T,\mu)=\frac{\pm L\sqrt{e^2 \left(4 \pi ^2 L^2 T^2-3\right)+12 \pi  G_N \mu ^2}+2
   \pi  e L^2 T}{3 e}\,.
\end{equation}
The physical branch is the positive branch because the $T\to 0$ limit of the negative branch is negative. However, the flat-space limit of the negative branch recovers \eqref{eqn:RN grand canonical horizon radius} correctly
\begin{equation}
    \lim_{L\to \infty}\frac{- L\sqrt{e^2 \left(4 \pi ^2 L^2 T^2-3\right)+12 \pi  G_N \mu ^2}+2\pi e L^2 T}{3 e}=\frac{e^2-4 \pi  G_N \mu ^2}{4 \pi  e^2 T} \, ,
\end{equation}
while the positive one does not.
Therefore special care needs to be taken and we should work directly with the RN grand canonical ensemble instead of naively extrapolating the RN-AdS results\footnote{Another way to see this is that the flat space RN BH descends from the small RN BH branch in AdS, that exists and is physical in the microcanonical ensemble (in the other ensembles it is unstable).}.

We start with the near-extremal one-loop corrected RN grand canonical partition function
\begin{equation}
\label{eqn:RN grand partition function from partition function}
    \mathcal{Z}_\text{RN}[\beta,\mu]=\sum_{Q\in e\cdot\mathbb{Z}}e^{\beta\mu\frac{Q}{e}} Z_\text{RN}[\beta,Q]=\sum_{Q\in e\cdot\mathbb{Z}}e^{\beta\mu\frac{Q}{e}-\beta M_0(Q)+S_0(Q)}Z_\text{Sch}(C(Q),\beta)\,.
\end{equation}
If we try to repeat the analysis as in the RN-AdS case, we find that the term in the exponent
\begin{equation}
\label{eqn:RN O(beta) no saddle point}
    \beta\mu \frac{Q}{e}-\beta M_0(Q)=\beta \left(\mu-\frac{e}{\sqrt{4\pi G_N}}\right)\frac{Q}{e} \, ,
\end{equation}
does not admit any stationary value of $Q$. Also, the RN grand canonical ensemble has the peculiar feature that at extremality, the chemical potential is frozen to
\begin{equation}
\label{eqn:RN extremal chemical potential}
    \mu=\frac{e}{\sqrt{4\pi G_N}}\,.
\end{equation}
To understand how we should proceed, we include here an interlude on the analysis of the BTZ BH. The grand canonical ensemble for the BTZ exhibits the same issues, but its one-loop grand canonical partition function is known and better understood \cite{Giombi:2008vd}. This provides an analytic playground for understanding how to adapt the RN-AdS analysis to the RN case.

The one-loop corrected BTZ grand canonical partition function is
\begin{equation}
    \mathcal{Z}_\text{BTZ}[\beta,\theta]=\chi_\mathds{1}(-1/\tau)\chi_\mathds{1}(1/\Bar{\tau})\,,
\end{equation}
where $\beta$ is the inverse temperature, $\theta=i\beta\Omega$ is the angular potential, $\tau=\frac{1}{2\pi}(\theta+i\beta)$ is the modular parameter of the torus (the conformal boundary of the BTZ), and
\begin{equation}
    \chi_\mathds{1}(\tau)=\frac{(1-q)q^{-\frac{c-1}{24}}}{\eta(\tau)},\quad q=e^{2\pi i\tau},\quad \eta(\tau)\text{: the Dedekind eta function}\,,
\end{equation}
is the Virasoro character of the vacuum representation. Setting the AdS$_3$ radius to be one, the central charge $c=3/2G_N$ to leading order in $G_N$. In the near-extremal regime \cite{Ghosh:2019rcj},
\begin{equation}
\label{eqn:exact near-extremal BTZ grand canonical partition function}
    \mathcal{Z}_\text{BTZ}[\beta,\theta]\approx 2\pi\left(\frac{\pi}{\beta}\right)^\frac{3}{2}\exp(\frac{c}{12}\frac{\pi^2}{\beta}+\frac{c}{6}\frac{\pi^2}{\beta}\frac{1}{1-\Omega}+\frac{\beta}{12})\,,
\end{equation}
\eqref{eqn:exact near-extremal BTZ grand canonical partition function} can be reproduced by the procedure described in this work. We refer the reader to \cite{Ghosh:2019rcj} for details. Starting from
\begin{equation}
\label{eqn:BTZ grand partition function from partition function}
    \mathcal{Z}_\text{BTZ}[\beta,\Omega]= \sum_{J\in\mathbb{Z}} e^{\beta\Omega J -\beta M_0(J)+S_0(J)}Z_\text{Sch}\left(\frac{c}{24},\beta\right) \, ,
\end{equation}
where $J$ is the rotational charge and
\begin{equation}
    M_0(J)=J,\quad S_0(J)=\pi\sqrt{\frac{J}{G_N}} \, ,
\end{equation}
we find an issue analogous to \eqref{eqn:RN O(beta) no saddle point}, namely that
\begin{equation}
\label{eqn:BTZ O(beta) no saddle point}
    \beta\Omega J-\beta M_0(J)=\beta(\Omega-1)J\,,
\end{equation}
has no stationary point in $J$. Moreover, for the BTZ BH,
\begin{equation}
    T=\frac{r_+^2-r_-^2}{2\pi r_+},\quad\Omega=\frac{r_-}{r_+} \, ,
\end{equation}
so that the angular velocity $\Omega$ is frozen to one at extremality, in analogy with \eqref{eqn:RN extremal chemical potential}. 

To obtain \eqref{eqn:exact near-extremal BTZ grand canonical partition function} from \eqref{eqn:BTZ grand partition function from partition function}, we essentially\footnote{One might notice that \eqref{eqn:exact near-extremal BTZ grand canonical partition function} contains an extra term $\frac{\beta}{12}$ in the exponent, which corresponds to a renormalization of the extremal energy $J\to J-\frac{1}{12}$. Also, \eqref{eqn:BTZ grand partition function from partition function} has an extra factor of $c^\frac{3}{2}$ due to $Z_\text{Sch}$. They are not important for our purposes, because these ‘mismatches’ already appear at the level of the near-extremal canonical partition function \cite{Ghosh:2019rcj}, and our goal here is to understand how to handle the sum over $J$ in \eqref{eqn:BTZ grand partition function from partition function}.} need
\begin{equation}
\label{eqn:extremal zero-loop BTZ grand canonical partition function}
    \sum_{J\in\mathbb{Z}} e^{\beta\Omega J -\beta M_0(J)+S_0(J)}\stackrel{!}{=}\exp(\frac{\pi^2}{4G_N\beta}\frac{1}{1-\Omega})\,.
\end{equation}
The right-hand side of \eqref{eqn:extremal zero-loop BTZ grand canonical partition function} is the zero-loop grand canonical partition function
\begin{equation}
    \mathcal{Z}_\text{BTZ}^\text{zero-loop}[\beta,\Omega]=e^{-\beta\Omega_\text{BTZ}(\beta,\Omega)}=\exp(\frac{\pi^2}{2G_N\beta}\frac{1}{1-\Omega^2})\,,
\end{equation}
in the extremal limit defined by $\beta\to\infty$ and $\Omega\to 1$, while keeping $\beta(1-\Omega)$ fixed. We now understand that due to the peculiar feature that $\Omega$ is frozen to one at extremality, the extremal zero-loop grand canonical partition function is not obtained by evaluating $e^{\beta\Omega J -\beta M_0(J)+S_0(J)}$ at the stationary point of $\beta\Omega J -\beta M_0(J)$, but rather at the stationary point of the complete exponent
\begin{equation}
\label{eqn:stationary J for BTZ}
    \left.\pdv{}{J}\left(\beta\Omega J-\beta M_0(J)+ S_0(J)\right)\right\rvert_{J=J_0}=0\quad\Rightarrow\quad J_0=\frac{\pi^2}{4G_N\beta^2}\frac{1}{(1-\Omega)^2}\,.
\end{equation}
Analogous to the RN-AdS analysis in \eqref{eqn:RN-AdS O(beta) saddle} and \eqref{eqn:mean value of the charge in the RN-AdS grand canonical ensemble}, $J_0$ corresponds to the extremal mean value of the angular momentum in the BTZ grand canonical ensemble
\begin{equation}
\label{eqn:mean value of the angular momentum in the BTZ grand canonical ensemble}
    J_0=\lim_{\substack{\beta\to\infty,\,\Omega\to 1,\\\beta(1-\Omega)\text{ fixed}}}-\pdv{}{\Omega}\Omega_\text{BTZ}(\beta,\Omega)=\frac{\pi^2}{4G_N\beta^2}\frac{1}{(1-\Omega)^2}\,.
\end{equation}
This stationary point is stable only when $\Omega <1$. In this case, we approximate the sum over $J$ by the stationary value $J_0$ and the result is \eqref{eqn:extremal zero-loop BTZ grand canonical partition function}. Thus, the one-loop correction is essentially reproduced by the Schwarzian action alone
\begin{equation}
    \mathcal{Z}_\text{BTZ}^\text{one-loop}[\beta,\Omega]=Z_\text{Sch}\left(\frac{c}{24},\beta\right)=\int\frac{\mathcal{D}\tau}{SL(2,\mathbb{R})}\,\exp(\frac{c}{24}\int_0^\beta\diff u\,\left\{\tan\frac{\pi\tau(u)}{\beta},u\right\})\,.
\end{equation}
In contrast with \eqref{eqn:1D effective theory for one-loop RN-AdS}, the contribution from the expected extremal zero modes associated with the $U(1)$ rotational isometry of the BTZ BH is absent.

To understand why, we follow the RN-AdS analysis and expand $J$ around $J_0$: $J=J_0+j$, where $j\in\mathbb{Z}$. To leading order in $j$,
\begin{equation}
    \beta\Omega J-\beta M_0(J)+ S_0(J)\approx\beta\Omega J_0-\beta M_0(J_0)+ S_0(J_0)-\frac{G_N\beta^3(1-\Omega)^3}{\pi^2}j^2\,.
\end{equation}
The expected extremal rotational zero modes contribute
\begin{equation}
\label{eqn:BTZ one-loop correction due to the would-be extremal rotational zero modes}
    \sum_{j\in\mathbb{Z}}e^{-\frac{G_N\beta^3(1-\Omega)^3}{\pi^2}j^2}=\sum_{j\in\mathbb{Z}}e^{-\beta\frac{j^2}{2K_\text{BTZ}}}=\int\frac{\mathcal{D}\theta}{U(1)}\,\exp(-\frac{K_\text{BTZ}}{2}\int_0^\beta\diff u\,\theta'(u)^2)\,,
\end{equation}
where
\begin{equation}
    K_\text{BTZ}=\lim_{\substack{\beta\to\infty,\,\Omega\to 1,\\\beta(1-\Omega)\text{ fixed}}}-\pdv[2]{}{\Omega}\Omega_\text{BTZ}(\beta,\Omega)=\frac{\pi^2}{2G_N\beta^3(1-\Omega)^3}\beta\to\infty \,.
\end{equation}
In the extremal limit, $K_\text{BTZ}$ diverges linearly with $\beta$ because $\beta(1-\Omega)$ is fixed. Since $K_\text{BTZ}$ appears together with $\beta^{-1}$ in \eqref{eqn:BTZ one-loop correction due to the would-be extremal rotational zero modes}, we conclude that the contribution from the expected extremal rotational zero modes is temperature-independent. In this work, we study the leading temperature-dependent quantum corrections and therefore we find that the rotational  zero modes do not contribute to leading order.

Our BTZ analysis is complementary to a criterion for identifying when an expected extremal gauge or rotational zero mode will be absent proposed in \cite{Kolanowski:2024zrq}: whenever an extremal gauge or rotational zero mode has a charge susceptibility
\begin{equation}
    K_\text{RN-AdS}=\frac{1}{e}\left.\pdv{Q}{\mu}\right\rvert_{T=0}=\Tilde{K},\quad K_\text{RN}=\frac{1}{e}\left.\pdv{Q}{\mu}\right\rvert_{T=0}=-\infty,\quad K_\text{BTZ}=\left.\pdv{J}{\Omega}\right\rvert_{T=0}=\infty\,,
\end{equation}
that diverges at extremality, the mode does not contribute to quantum corrections in the near-extremal regime. Since $K_\text{RN}$ diverges at extremality\footnote{The divergence of $K_\text{RN}$ directly follows from the fact that the chemical potential at extremality is frozen. The negative sign will be explained later.}, we expect that the extremal gauge zero modes do not contribute (to leading order) to the near-extremal one-loop RN grand canonical partition function. In other words, $\mathcal{Z}_\text{RN}^\text{one-loop}$ should essentially be reproduced by the Schwarzian action alone. 

We now extend the BTZ analysis to the RN case. The zero-loop RN grand canonical partition function is
\begin{equation}
    \mathcal{Z}_\text{RN}^\text{zero-loop}[\beta,\mu]=e^{-\beta\Omega_\text{RN}(\beta,\mu)}=\exp(-\frac{\beta^2(e^2-4\pi G_N \mu^2)^2}{16\pi G_N e^4})\,.
\end{equation}
In the extremal limit defined by $\beta\to \infty$ and $\mu\to e/\sqrt{4\pi G_N}$, while keeping $\beta(e-\mu\sqrt{4\pi G_N})$ fixed, it becomes
\begin{equation}
    \mathcal{Z}_\text{RN}^\text{zero-loop}[\beta,\mu]\to \exp(-\frac{\beta^2(e-\mu\sqrt{4\pi G_N})^2}{4\pi G_N e^2})=e^{\beta\mu \frac{Q_0}{e}-\beta M_0(Q_0)+S_0(Q_0)}\,,
\end{equation}
where $Q_0$ is simultaneously the extremal mean value of the charge
\begin{equation}
    Q_0=\lim_{\substack{\beta\to\infty,\,\mu\to e/\sqrt{4\pi G_N},\\\beta(e-\mu\sqrt{4\pi G_N})\text{ fixed}}}-e\pdv{}{\mu}\Omega_\text{RN}(\beta,\mu)=\frac{2\beta(e-\mu\sqrt{4\pi G_N})}{e\sqrt{4\pi G_N}}\,, 
\end{equation}
and the stationary point satisfying
\begin{equation}
    \left.\pdv{}{Q}\left(\beta\mu\frac{Q}{e}-\beta M_0(Q)+S_0(Q)\right)\right\rvert_{Q=Q_0}=0\quad \Rightarrow\quad Q_0=\frac{2\beta(e-\mu\sqrt{4\pi G_N})}{e\sqrt{4\pi G_N}}\,. 
\end{equation}
In accordance with the expectation that extremal gauge zero modes do not contribute, we do not further expand $Q$ around $Q_0$\footnote{If we expand $Q$ around $Q_0$ as $Q=Q_0+e\cdot q$ with $q\in\mathbb{Z}$, we obtain a divergent but temperature-independent sum: $\sum_{q\in\mathbb{Z}}e^{\frac{1}{4}e^2q^2}=\sum_{q\in\mathbb{Z}}e^{-\beta\frac{q^2}{2K_\text{RN}}}$, where 
\begin{equation}
    K_\text{RN}=\lim_{\substack{\beta\to\infty,\,\mu\to e/\sqrt{4\pi G_N},\\\beta(e-\mu\sqrt{4\pi G_N})\text{ fixed}}}-\pdv[2]{}{\mu}\Omega_\text{RN}(\beta,\mu)=-\frac{2\beta}{e^2}\to-\infty\,.
\end{equation}
The negativity of the charge susceptibility — or equivalently, the divergence of this sum — is a manifestation of the instability of the RN grand canonical ensemble in an asymptotically flat spacetime. We will not attempt to resolve this pathological feature here (see \cite{Braden:1990hw} for a resolution by placing the BH inside a cavity). Instead, we simply note that this contribution is temperature-independent and therefore does not affect the leading near-extremal quantum corrections.}, and we therefore propose that the near-extremal one-loop corrected RN grand canonical partition function is essentially described by
\begin{equation}
\label{eqn:1D effective theory for RN grand canonical}
    \mathcal{Z}_\text{RN}[\beta,\mu]=e^{\beta\mu \frac{Q_0}{e}-\beta M_0(Q_0)+S_0(Q_0)} \int\frac{\mathcal{D}\tau}{SL(2,\mathbb{R})}\,\exp(C(Q_0)\int_0^\beta\diff u\,\left\{\tan\frac{\pi\tau(u)}{\beta},u\right\})\,.
\end{equation}

\section{The RN BH interacting with a charged scalar field}
\label{sec:The RN BH Interacting with a Charged Scalar}

In this section, we send a wave of a charged-$e$ scalar field $\phi$ with mass $m$ into a near-extremal RN BH. We are interested to understand the properties of the quantum-corrected reflection coefficient and absorption cross-section of the BH.

Let $\phi$ be in a coherent state at frequency $\omega$ and with expected particle number $N_\omega$. We will assume the wavenumber $k=\sqrt{\omega^2-m^2}$ is small, so that $kr_+\ll 1$. This low-$k$ limit is essential because we will compute the absorption and emission rates using Fermi’s golden rule, which applies only when the interaction of the BH with the scalar field is weak. In our case, the relevant interaction strength is proportional to some positive power of $kr_+$. 

The core idea is to dimensionally reduce the 4D scalar\footnote{The factor of $r_+^{-1}$ is included to ensure that the kinetic terms for both $\phi$ and $\phi_{\ell m}$ are canonically normalized.}
\begin{equation}
    \phi(t,r,\Omega_2)=\frac{1}{r_+}\sum_{\ell,m}\phi_{\ell m}(t,r)Y_{\ell m}(\Omega_2)\,,
\end{equation}
that propagates on the near-extremal RN BH background to a 2D scalar $\phi_{\ell m}(t,r)$ that propagates in the near-horizon AdS$_2$ region by integrating out the $S^2$ degrees of freedom and zooming into $r-r_+\ll r_+$. In the low-$k$ limit, the dominant contribution to observables such as the Hawking emission rate and the absorption cross-section is provided by the s-wave sector $\ell=0$. Hence, for simplicity, we will focus on the s-wave component $\phi_{00}$ in what follows, although we will comment on the effects of higher partial waves at the end. Asymptotically, solutions to the scalar wave equation in AdS$_2$ take the form
\begin{equation}
\label{eqn:AdS2 scalar asymptotics}
    \phi_{00}(t,r)=\phi_\text{bdy}(t)r^{-1+\Delta}+O(r^{-\Delta})\,,
\end{equation}
where $\Delta$ is the scaling dimension of the boundary operator $\mathcal{O}$ dual to the AdS$_2$ scalar $\phi_{00}$. $\mathcal{O}$ is also an operator that exists in the 1D effective
theory \eqref{eqn:1D effective theory for RN grand canonical}. The coefficient $\phi_\text{bdy}(t)$ acts as a source for the operator $\mathcal{O}$ and deforms the 1D effective action by
\begin{equation}
    I_\text{int}=\int\diff t\,\phi_\text{bdy}(t)\mathcal{O}(t)\,.
\end{equation}

Let $c_{\omega \ell m}$ denote the modes of $\phi_{\ell m}(t,r)$ associated with the $O(r^{-1+\Delta})$ term in its expansion near the asymptotic AdS$_2$ boundary. Its precise definition will be given in \eqref{eqn:precise definition of the mode c}; for now, we simply note that
\begin{equation}
\label{eqn:mode expansion of phi_bdy in terms of c}
    \phi_\text{bdy}(t)\propto c_{\omega 00}e^{-i\omega t}\,.
\end{equation}
After quantizing the 4D scalar $\phi$, $\phi_\text{bdy}(t)$ becomes an operator acting on the matter Hilbert space $\mathcal{H}_\text{matter}$ at the asymptotic infinity. Let $\mathfrak{a}_{\omega\ell m}$ and $\mathfrak{a}_{\omega\ell m}^\dagger$ denote the annihilation and creation operators acting on $\mathcal{H}_\text{matter}$. They satisfy the canonical commutation relation
\begin{equation}
    [\mathfrak{a}_{\omega\ell m},\mathfrak{a}_{\omega'\ell'm'}^\dagger]=\delta(\omega-\omega')\delta_{\ell\ell'}\delta_{mm'}\,.
\end{equation}
In terms of $\mathfrak{a}_{\omega\ell m}$,
\begin{equation}
    \hat{\phi}_\text{bdy}(t)\propto \mathfrak{a}_{\omega 00}\frac{c_{\omega 00}}{a_{\omega 00}}e^{-i\omega t}\,,
\end{equation}
where $a_{\omega \ell m}$ are the modes  of $\phi$ associated with part of the asymptotic expansion at infinity of the ingoing wave $e^{-i\omega t-ir\sqrt{\omega^2-m^2}}$. Its precise definition will be given in \eqref{eqn:definition of canonically normalized a}.

The scalar field wave we send to the BH acts as an oscillating source at the boundary of the AdS$_2$ throat, exciting the BH. The response of the BH is encoded in the matrix elements of $\mathcal{O}$, which appears in the interaction Hamiltonian
\begin{equation}
\label{eqn:interaction Hamiltonian}
    H_\text{int}(t)=\hat{\phi}_\text{bdy}(t)\mathcal{O}(t)\,.
\end{equation}
This interaction drives transitions between BH states $\ket{E,Q}\in\mathcal{H}_\text{BH}$:
\begin{align}
    \ket{E_i,Q_i}\to\ket{E_i+\varpi,Q_i+e} & \quad \text{absorption}\\
    \ket{E_i,Q_i}\to\ket{E_i-\varpi,Q_i-e} & \quad \text{stimulated emission}\,,
\end{align}
where we have defined $\varpi=\omega-\mu$. The reason for using the variable $\varpi$ instead of $\omega$ is explained in appendix~\ref{subsec:The Density of States and Matter Correlators}.

$H_\text{int}$ acts on the tensor product of the the BH and matter Hilbert spaces $\mathcal{H}_\text{BH}\otimes\mathcal{H}_\text{matter}$. We adopt the notation $\ket{E,Q;N}=\ket{E,Q}\otimes\ket{N}$, where $Q$ is the BH charge, $E$ is the excitation energy of the BH above $M_0(Q)$, and $N$ is the occupation number for the scalar field's mode of energy $\omega$. We assume that the expected particle number of the wave is large, such that $\abs{N_f-N_i}\ll N_i\sim N_f \sim N_\omega$. Under this assumption, coherent states are well approximated by eigenstates of the occupation number. Also, spontaneous emission processes related to the Schwinger effect can be neglected (see~\cite{Brown:2024ajk} for an analysis of such processes).

The transition amplitudes between the initial and final states of this BH–matter system are given by Fermi’s golden rule
\begin{equation}
    \mathcal{T}_{i,f}=2\pi\rho(E_f,Q_f)\abs{\braket{E_f,Q_f;N_f|H_\text{int}|E_i,Q_i;N_i}}^2\,.
\end{equation}
The product of the BH final density of states and the matrix element of $\mathcal{O}$ is given by \eqref{eqn:density of states times matrix element}. The matrix element of $\hat{\phi}_\text{bdy}$ is
\begin{equation}
    \abs{\braket{N_f|\hat{\phi}_\text{bdy}|N_i}}^2=\mathcal{N}^2 \abs{\frac{c_{\omega 00}}{a_{\omega 00}}}^2 N_\omega\,,
\end{equation}
where $N_\omega=\abs{\braket{N_f|\mathfrak{a}_{\omega 00}|N_i}}^2$ is the expected particle number of the wave, and $\mathcal{N}^2$ is a normalization constant that will be determined later.

Taking into account a delta-function $\delta(E_f-E_i\pm\varpi)$, which enforces energy conservation, and integrating over the final energies, we can express the absorption and emission rates per unit frequency as
\begin{equation}
\label{eqn:absorption rates}
    \Gamma_\text{abs}(\omega)=2\pi\cdot \mathcal{N}^2 \abs{\frac{c_{\omega 00}}{a_{\omega 00}}}^2 N_\omega\cdot\rho(E_i+\varpi,Q_i+e)\abs{\braket{E_i+\varpi,Q_i+e|\mathcal{O}|E_i,Q_i}}^2\,,
\end{equation}
\begin{equation}
\label{eqn:emission rates}
    \Gamma_\text{emit}(\omega)=2\pi\cdot \mathcal{N}^2 \abs{\frac{c_{\omega 00}}{a_{\omega 00}}}^2 N_\omega\cdot\rho(E_i-\varpi,Q_i-e)\abs{\braket{E_i-\varpi,Q_i-e|\mathcal{O}|E_i,Q_i}}^2\,.
\end{equation}
To calculate the absorption coefficient, we have to subtract the stimulated emission \cite{Emparan:2025sao}
\begin{equation}
\label{eqn:definition of the absorption coefficient}
    P_\text{abs}(\omega)=\frac{2\pi}{N_\omega}\left(\Gamma_\text{abs}(\omega)-\Gamma_\text{emit}(\omega)\right)\,.
\end{equation}
The relation between the reflection coefficient and the absorption coefficient is
\begin{equation}
    \abs{\mathcal{R}}^2=1-P_\text{abs}(\omega)\,.
\end{equation}
The absorption cross-section is \cite{Benone:2015bst}
\begin{equation}
\label{eqn:quantum-corrected absorption cross-section}
\begin{split}
    \sigma_\text{abs}(\omega)=&\frac{\pi}{\omega^2-m^2}(1-\abs{\mathcal{R}}^2)\\
    =&\frac{4\pi^3\mathcal{N}^2}{\omega^2-m^2}\abs{\frac{c_{\omega 00}}{a_{\omega 00}}}^2\biggl(\rho(E_i+\varpi,Q_i+e)\abs{\braket{E_i+\varpi,Q_i+e|\mathcal{O}|E_i,Q_i}}^2\\
    &\hphantom{\frac{4\pi^3\mathcal{N}^2}{\omega^2-m^2}\abs{\frac{c_{\omega 00}}{a_{\omega 00}}}^2\biggl(}-\rho(E_i-\varpi,Q_i-e)\abs{\braket{E_i-\varpi,Q_i-e|\mathcal{O}|E_i,Q_i}}^2\biggl)\,.
\end{split}
\end{equation}
The only remaining task is to compute the ratio $\abs{c_{\omega00}/a_{\omega00}}^2$ and determine the normalization constant $\mathcal{N}^2$. 

\subsection{Computing the ratio $\abs{c_{\omega00}/a_{\omega00}}^2$ from the connection problem}
\label{subsec:Computing the Ratio abs{c_{omega00}/a_{omega00}}^2}

In order to compute the ratio $\abs{c_{\omega00}/a_{\omega00}}^2$, we have to solve the equation of motion for the charged scalar, and then expand the solution at asymptotic infinity and at the boundary of the AdS$_2$ throat respectively. We then need to relate these two solutions (by solving the so-called connection problem of the radial differential equation).

We first start with the equation of motion for the massive charged scalar on the RN BH background
\begin{equation}
\label{eqn:scalar EOM}
    g^{\mu\nu}(\nabla_\mu-ie A_\mu)(\nabla_\nu-ie A_\nu)\phi=m^2\phi \, ,
\end{equation}
where for notational simplicity, we will set $G_N=1$ and rescale the gauge coupling $e$ such that the RN metric (in Lorentzian signature) becomes
\begin{equation}
    g=-f(r)\diff t^2+\frac{\diff r^2}{f(r)}+r^2\diff\Omega_2^2,\quad f(r)=1-\frac{2M}{r}+\frac{Q^2}{r^2} \, .
\end{equation}
The gauge field, after a gauge transformation, becomes
\begin{equation}
    A=-\frac{Q}{r}\diff t \, .
\end{equation}
The separation of variables ansatz
\begin{equation}
    \phi(t,r,\Omega_2)=e^{-i\omega t}R(r)Y_{\ell m}(\Omega_2) \, ,
\end{equation}
reduces \eqref{eqn:scalar EOM} to the radial equation
\begin{equation}
\label{eqn:scalar radial equation}
    \dv{}{r} \left( r^2 f(r) \dv{R(r)}{r} \right)   +\left(\frac{( \omega + e A_t(r) )^2 r^2} {f(r)} - m^2 r^2 - \ell (\ell+1) \right) R(r)=0 \, .
\end{equation}
The radial ODE \eqref{eqn:scalar radial equation} turns out to be a confluent Heun equation, which does not admit a closed-form solution. Nevertheless, the exact connection formula for the expansions of $R(r)$ around $r=\infty$ and $r=r_+$ has been worked out \cite{Bonelli:2021uvf}. The key insight of \cite{Bonelli:2021uvf} is to observe that a certain BPZ equation that appears in 2D CFTs reduces to a confluent Heun equation in a corresponding semiclassical limit. Its solution, which can be thought of as a 4-point correlator involving the insertion of a certain degenerate field, admits conformal block expansions in the $t$ and $u$-channels. The crossing symmetry can be exploited to provide the connection formula between the two channels. A dictionary has been established in \cite{Bonelli:2021uvf}, such that the $t$-channel corresponds to the expansion of $R(r)$ around $r=r_+$ and the $u$-channel corresponds to the expansion of $R(r)$ around $r=\infty$. This dictionary in \cite{Bonelli:2021uvf} has been applied to study a charged massless scalar propagating on the Kerr-Newman BH background in \cite{Maulik:2025hax}. We refer the reader to \cite{Maulik:2025hax, Bonelli:2021uvf} for details. Here we will just present the results.

We define
\begin{equation}
    \Delta(r)=(r-r_+)(r-r_-),\quad z=\frac{r-r_-}{r_+-r_-},\quad \psi(z)=\Delta(r)^\frac{1}{2} R(r),\quad V(z)=\frac{1}{z^2(z-1)^2}\sum_{i=0}^4 A_i z^i\,,
\end{equation}
such that the radial equation \eqref{eqn:scalar radial equation} becomes a confluent Heun equation
\begin{equation}
\label{eqn:confluent Heun}
    \dv[2]{\psi(z)}{z}+V(z)\psi(z)=0\,,
\end{equation}
with 
\begin{equation}
    \begin{dcases}
    A_0=\frac{r_-^2 \left(e Q-r_- \omega \right)^2}{\left(r_--r_+\right)^2}+\frac{1}{4}\\
    A_1=\frac{2 r_- \left(r_- \omega -e Q\right) \left(2 r_- \omega -e
   Q\right)}{r_+-r_-}+\ell(\ell+1)+m^2 r_-^2\\
    A_2=e^2 Q^2+r_- \left(-6 e Q \omega -3 r_- \left(m^2-2 \omega ^2\right)+2 m^2 r_+\right)-\ell(\ell+1)\\
    A_3=\left(r_--r_+\right) \left(2 e Q \omega +r_- \left(3 m^2-4 \omega ^2\right)-m^2 r_+\right)\\
    A_4=\left(r_--r_+\right)^2 (\omega^2-m^2)\,.
\end{dcases}
\end{equation}
The BPZ equation studied in \cite{Bonelli:2021uvf} is parametrized by $a_1,a_2,m_3,\Lambda,E$, which is also a confluent Heun equation \eqref{eqn:confluent Heun} with
\begin{equation}
    \begin{dcases}
    A_0=\frac{1}{4}-a_1^2\\
    A_1=-\frac{1}{4}+E+a_1^2-a_2^2-m_3\Lambda\\
    A_2=\frac{1}{4}-E+2m_3\Lambda -\frac{\Lambda^2}{4}\\
    A_3=-m_3\Lambda+\frac{\Lambda^2}{2}\\
    A_4=-\frac{\Lambda^2}{4}\,.
\end{dcases}
\end{equation}
This leads to the dictionary between the BH problem and the BPZ problem
\begin{equation}
    \begin{dcases}
    \Lambda=-2i(r_+-r_-)\sqrt{\omega^2-m^2}\\
    m_3=\frac{i(eQ\omega+Mm^2-2M\omega^2)}{\sqrt{\omega^2-m^2}}\\
    E=-\omega ^2 \left(3 r_+^2+2 r_+ r_-+r_-^2\right)+r_+^2 m^2+2 e Q \omega  (2 r_++r_-)-e^2
    Q^2+\ell(\ell+1)+\frac{1}{4}\\
    a_1=\frac{ir_-(eQ-r_-\omega)}{r_+-r_-}\\
    a_2=\frac{ir_+(eQ-r_+\omega)}{r_+-r_-} \, .
\end{dcases}
\end{equation}
Notice that in the near-extremal limit that we are working $\abs{\Lambda}$ is a small parameter close to zero.

For the charged scalar propagating on the RN BH background, we impose an ingoing boundary condition at the horizon
\begin{equation}
    R(r\to r_+)\sim(r-r_+)^{-\frac{i(\omega- eQ/r_+)}{4\pi T}}=(r-r_+)^{a_2},\quad T=\frac{r_+-r_-}{4\pi r_+^2}\,.
\end{equation}
We then fix the normalization of our solution $\psi(z)=\Delta(r)^\frac{1}{2}R(r)$ to \eqref{eqn:confluent Heun}, such that as $z\to 1$ ($r\to r_+$),
\begin{equation}
\label{eqn:expansion of confluent Heun at horizon}
    \psi(z)=(z-1)^{\frac{1}{2}+a_2}\left(1+O(z-1)\right)\,.
\end{equation}
In this normalization, the use of crossing symmetry on the BPZ side provides us with the expansion of $R(r)$ around $r=\infty$ \cite{Bonelli:2021uvf}
\begin{equation}
\label{eqn:expansion of confluent Heun at infinity}
    \begin{split}
        \psi(z)=&\frac{\Gamma(1+2a)\Gamma(2a)\Gamma(1+2a_2)}{\Gamma(\frac{1}{2}+m_3+a)\Gamma(\frac{1}{2}\pm a_1+a_2+a)}\left.e^\frac{\Lambda}{2}\exp(\pdv{}{a_1}\mathcal{F}^\text{inst})\right\rvert_{a_1=a,\,a_2=-a}\\
        &\times e^\frac{\Lambda z}{2}\Lambda^{-\frac{1}{2}+m_3-a}z^{m_3}\left(1+O(z^{-1})\right)+(a\to -a)+\text{outgoing}\,,
    \end{split} 
\end{equation}
where we have suppressed the outgoing wave part and have included only the expression for the ingoing wave
\begin{equation}
    e^{\frac{\Lambda z}{2}}=e^{- i(r-r_-)\sqrt{\omega^2-m^2}} \, .
\end{equation}
The so-called instanton part of the Nekrasov-Shatashvili free energy is known perturbatively in $\Lambda$
\begin{equation}
    \mathcal{F}^\text{inst}=-\frac{\frac{1}{4}-a^2+a_1^2-a_2^2}{\frac{1}{2}-2a^2}m_3\Lambda +O(\Lambda^2)\,,
\end{equation}
where $a$ is defined by the so-called Matone’s relation
\begin{equation}
    E=a^2-\Lambda\pdv{}{\Lambda}\mathcal{F}^\text{inst}\quad \Rightarrow \quad a(E)=\sqrt{E}-\frac{\frac{1}{4}-E+a_1^2-a_2^2}{\sqrt{E}(1-4E)}m_3\Lambda +O(\Lambda^2)\,.
\end{equation}
From \eqref{eqn:expansion of confluent Heun at horizon} and \eqref{eqn:expansion of confluent Heun at infinity}, we can compute the greybody factor $P(\omega,\ell)$ defined as the ratio of ingoing fluxes ($\Im \psi^\dagger \partial_z \psi$ for $z\in\mathbb{R}$) at the horizon and 
asymptotic infinity
\begin{equation}
\begin{split}
    &P(\omega,\ell)=\frac{\omega-eQ/r_+}{2\pi T}\\
    &\times \abs{\frac{\Gamma(1+2a)\Gamma(2a)\Gamma(1+2a_2)}{\Gamma(\frac{1}{2}+m_3+a) \Gamma(\frac{1}{2}\pm a_1+a_2+a)}\left.e^\frac{\Lambda}{2}\exp(\pdv{}{a_1}\mathcal{F}^\text{inst})\right\rvert_{a_1=a,\,a_2=-a}\Lambda^{m_3-a}+(a\to -a)}^{-2}\,.
\end{split}
\end{equation}
The greybody factor appears both in classical scattering (where it serves as the classical analogue of $P_\text{abs}$ defined in \eqref{eqn:definition of the absorption coefficient}) and in the semiclassical Hawking emission rate. It will be used later to determine the normalization constant $\mathcal{N}^2$ by matching the quantum-corrected Hawking emission rate with the known semiclassical Hawking emission rate in the corresponding limit.

For now, let us proceed to work out the expansion of $R(r)$ at the boundary of the near-horizon AdS$_2$ throat. We remind the reader that we want to work out the ratio $\abs{c_{\omega00}/a_{\omega00}}^2$, which requires the expansion of $R(r)$ at the asymptotic infinity and at the boundary of the AdS$_2$ throat. We have already worked out the expansion of $R(r)$ at the asymptotic infinity when the normalization of $R(r)$ at the horizon is given by \eqref{eqn:expansion of confluent Heun at horizon}. Therefore, we only have to solve the radial equation \eqref{eqn:scalar radial equation} in the near-horizon region defined by 
\begin{equation}
    \frac{r-r_+}{r_+}\ll 1 \, ,
\end{equation}
using the same normalization. In the near-horizon approximation, the radial equation can be solved in a closed form. The expansion of this solution at $r\to\infty$, which corresponds to the asymptotic boundary of the near-horizon AdS$_2$ throat, is what we are after.

We define
\begin{equation}
    x=\frac{r-r_+}{r_+-r_-}\,,
\end{equation}
so that the radial equation becomes
\begin{equation}
    x(x+1)R''(x)+(2x+1)R'(x)+\left(-\ell(\ell+1)+\frac{p_4x^4+p_3x^3+p_2x^2+p_1x+p_0}{x(x+1)}\right)R(x)=0\,,
\end{equation}
where
\begin{equation}
    \begin{dcases}
    p_0=\frac{r_+^2 (e Q-r_+ \omega )^2}{(r_+-r_-)^2}\\
    p_1=\frac{r_+\left(2e^2Q^2+(m^2r_--6eQ\omega)r_++(4\omega^2-m^2)r_+^2\right)}{r_+-r_-}\\
    p_2=e^2Q^2+(2m^2r_--6eQ\omega)r_+ +3(2\omega^2-m^2)r_+^2\\
    p_3=(r_+-r_-)\left(m^2r_- -2eQ\omega+(4\omega^2-3m^2)r_+\right)\\
    p_4=(r_+-r_-)^2(\omega^2-m^2)\,.
\end{dcases}
\end{equation}
In the near-horizon region $(x\ll 1)$, we can approximate
\begin{equation}
    x(x+1)R''(x)+(2x+1)R'(x)+\left(-\ell(\ell+1)+\frac{p_2x^2+p_1x+p_0}{x(x+1)}\right)R(x)=0\,.
\end{equation}
The solution is 
\begin{equation}
    \begin{split}
        R(x)=&A_1 x^{-i\sqrt{p_0}}(x+1)^{i\sqrt{p_2-p_1+p_0}} \\
        &\times {}_2F_1\left(1-i\sqrt{p_0}+i\sqrt{p_2-p_1+p_0}-\Delta,-i\sqrt{p_0}+i\sqrt{p_2-p_1+p_0}+\Delta;1-2i\sqrt{p_0};-x\right)\\
        &+A_2 x^{i\sqrt{p_0}}(x+1)^{-i\sqrt{p_2-p_1+p_0}} \\
        &\times {}_2F_1\left(1+i\sqrt{p_0}-i\sqrt{p_2-p_1+p_0}-\Delta,i\sqrt{p_0}-i\sqrt{p_2-p_1+p_0}+\Delta;1+2i\sqrt{p_0};-x\right)\,,
    \end{split}
\end{equation}
where 
\begin{equation}
    \label{eqn:almost the AdS2 scalar scaling dimension}
    \Delta=\frac{1}{2}+\sqrt{\left(\frac{1+2\ell}{2}\right)^2-p_2}\,.
\end{equation}

Suppose $r_+\omega>eQ$. Then
\begin{equation}
    \sqrt{p_0}=\frac{r_+(r_+\omega-eQ)}{r_+-r_-}\,,
\end{equation}
and so the $A_1$ branch satisfies an ingoing boundary condition at the horizon. On the other hand if we suppose that $r_+\omega<eQ$, then the $A_2$ branch satisfies an ingoing boundary condition at the horizon. To proceed, we assume $r_+\omega>eQ$ and set $A_2=0$. Recall that we have chosen the normalization
\begin{equation}
    \psi(z\to 1)=\sqrt{(r-r_+)(r_+-r_-)}R(r\to r_+)=\left(\frac{r-r_+}{r_+-r_-}\right)^{\frac{1}{2}-\frac{i(\omega-eQ/r_+)}{4\pi T}}+\cdots\,,
\end{equation}
which implies
\begin{equation}
    R(x\to 0)=A_1\left(\frac{r-r_+}{r_+-r_-}\right)^{-\frac{i(\omega-eQ/r_+)}{4\pi T}}+\cdots \stackrel{!}{=}\frac{1}{r_+-r_-}\left(\frac{r-r_+}{r_+-r_-}\right)^{-\frac{i(\omega-eQ/r_+)}{4\pi T}}+\cdots\,.
\end{equation}
This fixes $A_1=(r_+-r_-)^{-1}$.

Near the asymptotic boundary of the near-horizon AdS$_2$ throat, we find
\begin{equation}
\begin{split}
\label{eqn:expansion of confluent Heun at AdS2 boundary}
    R(x\to\infty)=&\frac{x^{-1+\Delta}}{r_+-r_-}\frac{\Gamma(1-2i\sqrt{p_0})\Gamma(-1+2\Delta)}{\Gamma(-i\sqrt{p_0}\pm i\sqrt{p_2-p_1+p_0}+\Delta)}\\
    &+\frac{x^{-\Delta}}{r_+-r_-}\frac{\Gamma(1-2i\sqrt{p_0})\Gamma(1-2\Delta)}{\Gamma(1-i\sqrt{p_0}\pm i\sqrt{p_2-p_1+p_0}-\Delta)}\,.
\end{split}
\end{equation}
Note that this is precisely the expected asymptotic behavior \eqref{eqn:AdS2 scalar asymptotics} for an AdS$_2$ scalar. After taking the low-$k$ limit\footnote{Without taking the low-$k$ limit, \eqref{eqn:almost the AdS2 scalar scaling dimension} depends on $\omega$, which is unphysical because the scaling dimension $\Delta$ should depend only on the BH background parameters $Q,r_+$ and the parameters $m,e,\ell$ that characterize the AdS$_2$ scalar $\phi_{\ell m}$. Taking the low-$k$ limit is also necessary because the BPZ method solves the connection problem without assuming $kr_+\ll 1$.

In a more traditional treatment of this connection problem, we divide the BH exterior into several regions, and solve the radial equation in each region separately. For this method to work, these regions must have overlaps, and the low-$k$ condition is crucial for this to happen. For example, see \cite{Alberti:2025mpg} for an application of this method to a charged massless scalar on the near-extremal RN BH background.}, we can identify
\begin{equation}
    \label{eqn:AdS2 scalar scaling dimension}
    \Delta=\frac{1}{2}+\sqrt{\left(\frac{1+2\ell}{2}\right)^2-e^2Q^2+6eQmr_+-5m^2r_+^2}\,,
\end{equation}
as the scaling dimension of the boundary dual $\mathcal{O}$ of the AdS$_2$ scalar $\phi_{\ell m}$. Note that in the near-extremal regime $r_+\approx r_-$, we have $\abs{\Lambda}\to 0$. To leading order in $\Lambda$,
\begin{equation}
\label{eqn:BF bound}
    \Delta\in\mathbb{R}\quad \Leftrightarrow\quad  \left(\frac{1+2\ell}{2}\right)^2-e^2Q^2+6eQmr_+-5m^2r_+^2\approx E \approx a^2\geq 0\,,
\end{equation}
\eqref{eqn:BF bound} can be interpreted as the Breitenlohner–Freedman (BF) bound for $\phi_{\ell m}$. In this work, we focus on the case in which the BF bound is satisfied\footnote{It is also interesting to study the role of quantum corrections, in the case where the BF bound is violated, and in particular in order to elucidate their role in the instability of the charged scalar to condense in the near-horizon region.}.

We now have all the ingredients to compute the ratio $\abs{c_{\omega00}/a_{\omega00}}^2$. The mode $c_{\omega\ell m}$ is defined by the $O(r^{-1+\Delta})$ term in \eqref{eqn:expansion of confluent Heun at AdS2 boundary}
\begin{equation}
\label{eqn:precise definition of the mode c}
    c_{\omega\ell m}\left(\frac{r_+^2}{r-r_+}\right)^{1-\Delta}=\frac{1}{r_+-r_-}\left(\frac{r-r_+}{r_+-r_-}\right)^{-1+\Delta}\frac{\Gamma(1-2i\sqrt{p_0})\Gamma(-1+2\Delta)}{\Gamma(-i\sqrt{p_0}\pm i\sqrt{p_2-p_1+p_0}+\Delta)}\,,
\end{equation}
to be
\begin{equation}
    c_{\omega \ell m}=r_+^{2(\Delta-1)}(r_+-r_-)^{-\Delta}\frac{\Gamma(1-2i\sqrt{p_0})\Gamma(-1+2\Delta)}{\Gamma(-i\sqrt{p_0}\pm i\sqrt{p_2-p_1+p_0}+\Delta)}\,.
\end{equation}
In the near-extremal regime, the ingoing wave at asymptotic infinity can be approximated by
\begin{equation}
\label{eqn:near-extremal expansion of confluent Heun at asymptotic infinity}
    R(r)\approx \frac{\Gamma(1+2a)\Gamma(2a)\Gamma(1+2a_2)}{\Gamma(\frac{1}{2}+m_3+a)\Gamma(\frac{1}{2}\pm a_1+a_2+a)}\Lambda^{-\frac{1}{2}+m_3-a}\left(\frac{r}{r_+-r_-}\right)^{m_3}\frac{e^{-ir\sqrt{\omega^2-m^2}}}{r}\,.
\end{equation}
Here we have used the fact that the BF bound holds, implying $a>0$. Together with $\abs{\Lambda}\to 0$, this allows us to neglect the $(a\to -a)$ term in \eqref{eqn:expansion of confluent Heun at infinity}. The canonically normalized mode $a_{\omega\ell m}$ is defined by 
\begin{equation}
\label{eqn:definition of canonically normalized a}
    R(r)=\frac{r_+}{\sqrt{4\pi\sqrt{\omega^2-m^2}}} a_{\omega \ell m}\frac{e^{-ir\sqrt{\omega^2-m^2}}r^{m_3}}{r}\,.
\end{equation}
Matching \eqref{eqn:near-extremal expansion of confluent Heun at asymptotic infinity} with \eqref{eqn:definition of canonically normalized a} gives
\begin{equation}
    a_{\omega\ell m}=\frac{\sqrt{4\pi\sqrt{\omega^2-m^2}}}{r_+}\frac{\Gamma(1+2a)\Gamma(2a)\Gamma(1+2a_2)}{\Gamma(\frac{1}{2}+m_3+a)\Gamma(\frac{1}{2}\pm a_1+a_2+a)}\Lambda^{-\frac{1}{2}+m_3-a}\left(\frac{1}{r_+-r_-}\right)^{m_3}\,.
\end{equation}
Note that in the near-extremal regime,
\begin{equation}
    \Delta\approx \frac{1}{2}+a,\quad a_2=-i\sqrt{p_0},\quad a_1\approx -i\sqrt{p_2-p_1+p_0}\,,
\end{equation}
so that the ratio simplifies significantly
\begin{equation}
\label{eqn:ratio of c and a}
    \abs{\frac{c_{\omega\ell m}}{a_{\omega\ell m}}}^2\approx \frac{2^{-1+2a}r_+^{4a}(\omega^2-m^2)^a\abs{\Gamma(\frac{1}{2}+m_3+a)}^2e^{i\pi m_3}}{\pi\Gamma(1+2a)^2}\,.
\end{equation}
From \eqref{eqn:ratio of c and a}, we see that
\begin{equation}
    H_\text{int}(t)= g \, e^{-i\omega t} \mathfrak{a}_{\omega 00} \mathcal{O}(t) \propto \frac{c_{\omega00}}{a_{\omega 00}}e^{-i\omega t}\mathfrak{a}_{\omega 00} \mathcal{O}(t) \,  ,
\end{equation}
and so the interaction coupling $g$ is proportional to $(kr_+)^a$. It is therefore \textit{a posteriori} clear that assuming $kr_+\ll 1$ is crucial for the validity of applying Fermi’s golden rule.

 The final piece remaining is to fix the normalization constant $\mathcal{N}^2$, and we do so in the next section.

\subsection{Determining the overall normalization $\mathcal{N}^2$}
\label{subsec:Determining the Normalization N^2}

The semiclassical Hawking emission rate is
\begin{equation}
\label{eqn:semiclassical Hawking emission rate}
    \frac{\diff^2 N}{\diff t\diff\omega}=\frac{1}{2\pi}\,\frac{P(\omega,\ell)}{e^{\beta(\omega-eQ/r_+)}-1}\,,
\end{equation}
where the greybody factor $P(\omega,\ell)$ in the near-extremal limit reduces to
\begin{equation}
\label{eqn:semiclassical greybody factor}
    P(\omega,\ell)\approx \frac{\omega-eQ/r_+}{2\pi T}\abs{\frac{\Gamma(\Delta_\ell+m_3)\Gamma(\Delta_\ell)\Gamma(\Delta_\ell+ 2a_2)}{\Gamma(2\Delta_\ell)\Gamma(2\Delta_\ell-1)\Gamma(1+2a_2)}}^2 e^{i\pi m_3}(8\pi T kr_+^2)^{2\Delta_\ell-1} \, ,
\end{equation}
For the reader’s convenience, we collect the relevant definitions here
\begin{equation}
    \begin{dcases}
        \Delta_\ell=\frac{1}{2}+\sqrt{\left(\frac{1+2\ell}{2}\right)^2-e^2Q^2+6eQmr_+-5m^2r_+^2}\\
        m_3=\frac{i(eQ\omega-Mm^2)}{k}-2ikM\\
        a_2=\frac{i(eQ/r_+-\omega)}{4\pi T}\,.
    \end{dcases}
\end{equation}
We will mainly focus on the s-wave sector ($\ell=0$) and comment on the contributions from higher partial waves ($\ell>0$) at the end. Accordingly, we suppress the $\ell=0$ subscript of $\Delta_\ell$ when discussing the s-wave sector.

The quantum-corrected spontaneous emission rate given by the Fermi's golden rule 
\begin{equation}
\label{eqn:quantum-corrected Hawking emission rate}
\begin{split}
    &\frac{\diff^2 N}{\diff t\diff\omega}=2\pi\cdot \mathcal{N}^2 \abs{\frac{c_{\omega00}}{a_{\omega00}}}^2 \cdot\rho(E_i-\varpi,Q_i-e)\abs{\braket{E_i-\varpi,Q_i-e|\mathcal{O}|E_i,Q_i}}^2\\
    =&2\pi\cdot \mathcal{N}^2 \frac{2^{2(\Delta-1)}(kr_+^2)^{2\Delta-1}\abs{\Gamma(\Delta+m_3)}^2e^{i\pi m_3}}{\pi\Gamma(2\Delta)^2} \cdot \frac{C(Q_0)}{\pi^2}\sinh(2\pi\sqrt{2C(Q_0)(E_i-\varpi)})\\
        &\times \frac{\Gamma\left(\Delta\pm i\sqrt{2C(Q_0)(E_i-\varpi)}\pm i\sqrt{2C(Q_0)E_i}\right)}{(2C(Q_0))^{2\Delta}\Gamma(2\Delta)} \, .
\end{split} 
\end{equation}

We now fix $\mathcal{N}^2$ by matching the semiclassical expression \eqref{eqn:semiclassical Hawking emission rate} to the quantum-corrected result \eqref{eqn:quantum-corrected Hawking emission rate} in the semiclassical limit
\begin{equation}
\label{eqn:semiclassical limit}
    C(Q_0)E_i\to\infty\text{, while keeping $\frac{C(Q_0)}{E_i}$ fixed} \,.
\end{equation}
To understand why the semiclassical limit is given by \eqref{eqn:semiclassical limit}, we recall that $1/C(Q_0)$ is the energy scale below which the semiclassical thermodynamics breaks down. Hence, $C(Q_0)E_i\to\infty$ means that the initial energy excitation of the BH is far above the break-down scale $1/C(Q_0)$ and so, we are in the semiclassical regime. It will turn out that fixing $C(Q_0)/E_i$ also fixes the temperature of the BH.

Let us first assume that $\Delta = 1$. In this case, we use the identity
\begin{equation}
    \Gamma(1\pm iA\pm iB)=\frac{2\pi^2(A^2-B^2)}{\cosh(2\pi A)-\cosh(2\pi B)} \, ,
\end{equation}
to find
\begin{equation}
    \begin{split}
        &\frac{C(Q_0)}{\pi^2}\sinh(2\pi\sqrt{2C(Q_0)(E_i-\varpi)})\frac{\Gamma\left(1\pm i\sqrt{2C(Q_0)(E_i-\varpi)}\pm i\sqrt{2C(Q_0)E_i}\right)}{(2C(Q_0))^{2}}\\
        =&\frac{\varpi \sinh(2\pi\sqrt{2C(Q_0)(E_i-\varpi)})}{\cosh(2\pi\sqrt{2C(Q_0)E_i})-\cosh(2\pi\sqrt{2C(Q_0)(E_i-\varpi)})}\\
        \to&\frac{\varpi}{\exp(2\pi\sqrt{2C(Q_0)E_i}-2\pi\sqrt{2C(Q_0)(E_i-\varpi)})-1}\\
        \approx&\frac{\varpi}{e^{\beta\varpi}-1} \, ,
    \end{split}
\end{equation}
where in the last line we have assumed $\abs{\varpi}\ll E_i$ and identified the inverse BH temperature
\begin{equation}
    \beta=\sqrt{\frac{2\pi^2 C(Q_0)}{E_i}} \, ,
\end{equation}
as advertised.

In general $\Delta\neq 1$, so we need to take the same limit using the following factor in the emission rate 
\begin{equation}
    \frac{1}{(2C(Q_0))^{2(\Delta-1)}\Gamma(2\Delta)}\frac{\Gamma\left(\Delta\pm i\sqrt{2C(Q_0)(E_i-\varpi)}\pm i\sqrt{2C(Q_0)E_i}\right)}{\Gamma\left(1\pm i\sqrt{2C(Q_0)(E_i-\varpi)}\pm i\sqrt{2C(Q_0)E_i}\right)} \, .
\end{equation}
Since the chemical potential in our units is $\mu=eQ/r_+$, we note that 
\begin{equation}
    \abs{\Gamma\left(\Delta+ i\sqrt{2C(Q_0)(E_i-\varpi)}- i\sqrt{2C(Q_0)E_i}\right)}^2\approx \abs{\Gamma(\Delta+2a_2)}^2\,,
\end{equation}
and
\begin{equation}
    \abs{\Gamma\left(1+ i\sqrt{2C(Q_0)(E_i-\varpi)}- i\sqrt{2C(Q_0)E_i}\right)}^2\approx \abs{\Gamma(1+2a_2)}^2 \, ,
\end{equation}
and finally
\begin{equation}
    \abs{\frac{\Gamma\left(\Delta+ i\sqrt{2C(Q_0)(E_i-\varpi)}+ i\sqrt{2C(Q_0)E_i}\right)}{\Gamma\left(1+ i\sqrt{2C(Q_0)(E_i-\varpi)}+ i\sqrt{2C(Q_0)E_i}\right)}}^2\to\left(2\sqrt{2C(Q_0)E_i}\right)^{2(\Delta-1)}\,.
\end{equation}
The last limit follows from $\abs{\varpi}\ll E_i$, $C(Q_0)E_i\to\infty$, and
\begin{equation}
    \frac{\Gamma(n+ix)\Gamma(n-ix)}{\Gamma(1+ix)\Gamma(1-ix)}=x^{2(n-1)}+O(x^{2(n-2)}) \, .
\end{equation}
In total, we find
\begin{equation}
\label{eqn:semiclassical limit of DOS times matrix element}
    \begin{split}
        &\frac{C(Q_0)}{\pi^2}\sinh(2\pi\sqrt{2C(Q_0)(E_i-\varpi)}) \frac{\Gamma\left(\Delta\pm i\sqrt{2C(Q_0)(E_i-\varpi)}\pm i\sqrt{2C(Q_0)E_i}\right)}{(2C(Q_0))^{2\Delta}\Gamma(2\Delta)}\\
        \to&\frac{\varpi}{e^{\beta\varpi}-1}\frac{1}{\Gamma(2\Delta)}\abs{\frac{\Gamma(\Delta+2a_2)}{\Gamma(1+2a_2)}}^2\left(2\pi T\right)^{2(\Delta-1)} \, .
    \end{split}
\end{equation}
If we multiply \eqref{eqn:semiclassical limit of DOS times matrix element} by $2\pi \mathcal{N}^2\abs{c_{\omega00}/a_{\omega00}}^2$, where $\abs{c_{\omega00}/a_{\omega00}}^2$ is given by \eqref{eqn:ratio of c and a}, and then match with the semiclassical Hawking emission rate \eqref{eqn:semiclassical Hawking emission rate}, we find
\begin{equation}
    \mathcal{N}^2=\frac{2^{2(\Delta-1)}}{\pi}\frac{\Gamma(2\Delta)\Gamma(\Delta)^2}{\Gamma(2\Delta-1)^2}\,,
\end{equation}
or equivalently,
\begin{equation}
\label{eqn:normalization constant result}
    \mathcal{N}^2\abs{\frac{c_{\omega00}}{a_{\omega00}}}^2=\frac{2^{4(\Delta-1)}}{\pi^2}\frac{\Gamma(\Delta)^2\abs{\Gamma(\Delta+m_3)}^2}{\Gamma(2\Delta-1)^2\Gamma(2\Delta)}(k r_+^2)^{2\Delta-1}e^{i\pi m_3}\,.
\end{equation}
As a first sanity check, if our scalar is massless and neutral, then \eqref{eqn:quantum-corrected Hawking emission rate} becomes
\begin{equation}
    \frac{\diff^2 N}{\diff t\diff\omega}=\frac{2}{\pi}r_+^2\omega^2\frac{\sinh(2\pi\sqrt{2C(E_i-\omega)})}{\cosh(2\pi\sqrt{2CE_i})-\cosh(2\pi\sqrt{2C(E_i-\omega)})}\,,
\end{equation}
which agrees with the quantum-corrected Hawking emission rate for neutral scalars found in \cite{Brown:2024ajk}.

As another sanity check, in the small charge $eQ\ll 1$ and small mass $mr_+\ll 1$ limit, we have $\Delta\approx 1$ and so \eqref{eqn:quantum-corrected absorption cross-section} becomes
\begin{equation}
\label{eqn:semiclassical l=0 absorption cross-section}
    \sigma_\text{abs}(\omega)\to 4\pi r_+^2\frac{\varpi}{\sqrt{\omega^2-m^2}}\frac{\pi \abs{m_3}}{\sinh(\pi \abs{m_3})}e^{i\pi m_3}\,,
\end{equation}
in the semiclassical limit
\begin{equation}
    \rho(E_i+\varpi,Q_i+e)\abs{\braket{E_i+\varpi,Q_i+e|\mathcal{O}|E_i,Q_i}}^2\to \frac{\varpi e^{\beta\varpi}}{e^{\beta\varpi}-1}\quad (\Delta=1)\,,
\end{equation}
\begin{equation}
    \rho(E_i-\varpi,Q_i-e)\abs{\braket{E_i-\varpi,Q_i-e|\mathcal{O}|E_i,Q_i}}^2\to \frac{\varpi}{e^{\beta\varpi}-1}\quad (\Delta=1)\,,
\end{equation}
\eqref{eqn:semiclassical l=0 absorption cross-section} is the classical s-wave absorption cross-section for a charged massive scalar field computed in \cite{Gibbons:1975kk}.

\subsection{The reflection coefficient $\abs{\mathcal{R}}^2$ and absorption cross-section $\sigma_\text{abs}$}
\label{subsec:Plots of abs{R}^2 and sigma_{abs}}

In this section, we present and analyze a few plots comparing the quantum-corrected reflection coefficient and absorption cross-section with their semiclassical counterparts.

\begin{figure}[htbp]
    \centering
    \begin{subfigure}[b]{0.49\textwidth}
        \centering
        \includegraphics[width=\textwidth]{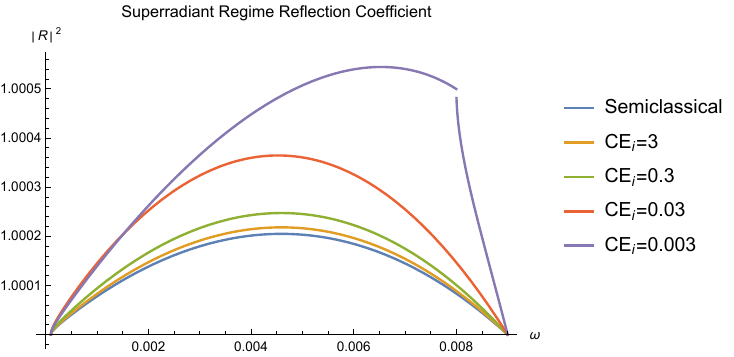}
        \caption{$\abs{\mathcal{R}}^2$ over the superradiant regime $[m,\mu)$.}
    \end{subfigure}
    \hfill
    \begin{subfigure}[b]{0.49\textwidth}
        \centering
        \includegraphics[width=\textwidth]{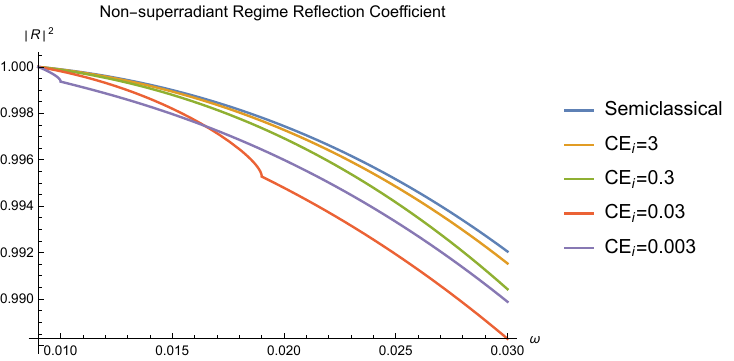}
        \caption{$\abs{\mathcal{R}}^2$ outside the superradiant regime.}
    \end{subfigure}
    \caption{The reflection coefficient is given by $\abs{\mathcal{R}}^2 = 1 - P_\text{abs}(\omega)$. The blue curve shows the semiclassical result, where $P_\text{abs}(\omega) = P(\omega, \ell=0)$ as defined in \eqref{eqn:semiclassical greybody factor}. The other curves show the quantum-corrected results for various values of $C(Q_0)E_i$, with $P_\text{abs}(\omega)$ given by \eqref{eqn:definition of the absorption coefficient}. The kinks within the superradiant regime appear at $\mu-E_i$ and correspond to the transition where $\Gamma_\text{abs}=0$ switches to $\Gamma_\text{abs}>0$. The kinks outside the superradiant region occur at $\mu+E_i$ and correspond to the transition where $\Gamma_\text{emit}>0$ switches to $\Gamma_\text{emit}=0$.}
    \label{fig:reflection coefficient}
\end{figure}

\begin{figure}[htbp]
    \centering
    \begin{subfigure}[b]{0.49\textwidth}
        \centering
        \includegraphics[width=\textwidth]{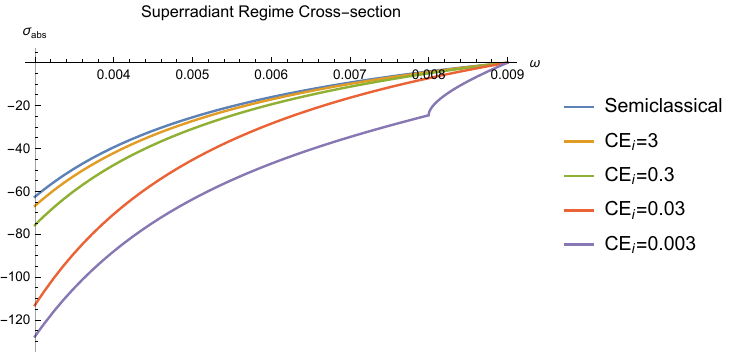}
        \caption{$\sigma_\text{abs}$ over the superradiant regime $[3\times10^{-3},\mu)$.}
    \end{subfigure}
    \hfill
    \begin{subfigure}[b]{0.49\textwidth}
        \centering
        \includegraphics[width=\textwidth]{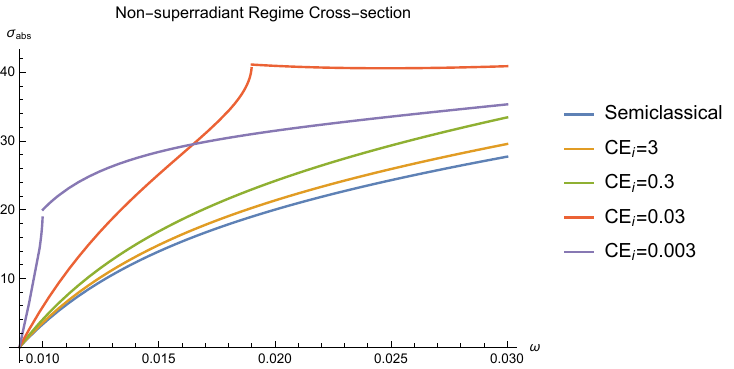}
        \caption{$\sigma_\text{abs}$ outside the superradiant regime.}
    \end{subfigure}
    \caption{The absorption cross-section $\sigma_\text{abs}$. The blue curve shows the semiclassical result, defined by \eqref{eqn:semiclassical l=0 absorption cross-section}. The other curves show the quantum-corrected results for different values of $C(Q_0)E_i$, computed using \eqref{eqn:quantum-corrected absorption cross-section}. The kinks within the superradiant regime appear at $\mu-E_i$ and correspond to the transition where $\Gamma_\text{abs}=0$ switches to $\Gamma_\text{abs}>0$. The kinks outside the superradiant region occur at $\mu+E_i$ and correspond to the transition where $\Gamma_\text{emit}>0$ switches to $\Gamma_\text{emit}=0$.}
    \label{fig:cross section}
\end{figure}

Specifically, in figures~\ref{fig:reflection coefficient} and~\ref{fig:cross section} we plot $\abs{\mathcal{R}}^2$ and $\sigma_\text{abs}$ as functions of $\omega$. The frequency range $[m,\omega_\text{max}]$ is divided into two intervals: $[m,\mu)$ and $(\mu,\omega_\text{max}]$. The first interval, $[m,\mu)$, corresponds to the superradiant regime \cite{Brito:2015oca}, where $\abs{\mathcal{R}}^2>1$ and $\sigma_\text{abs}<0$. The upper limit $\omega_\text{max}$ is chosen so that $r_+\sqrt{\omega_\text{max}^2-m^2}\ll1$, ensuring the validity of the low-$k$ assumption that underpins our analysis.

For the charged massive scalar, we choose in our plots the charge and mass to be $e=10^{-2}$ and $m=10^{-4}$ respectively. In our units, the relevant BH parameters are given by
\begin{equation}
    Q_0=\lim_{\substack{\beta\to\infty,\,\mu\to e,\\\beta(e-\mu)\text{ fixed}}}-e\pdv{}{\mu}\Omega_\text{RN}(\beta,\mu)=\frac{\mu}{e}\left(1-\frac{\mu}{e}\right)\frac{\beta}{2\pi},\quad r_+\approx \frac{e}{\mu}Q_0,\quad M\approx\frac{Q_0^2+r_+^2}{2r_+}\,,
\end{equation}
where we have used the approximation $Q\approx Q_0$, consistent with our assumption in appendix~\ref{subsec:The Density of States and Matter Correlators} that the background BH charge is much larger than any charge fluctuations. For the numerical evaluation, we set the inverse temperature to $\beta=10^2$ and the chemical potential to $\mu/e=0.9$. This choice ensures that $Q_0/e\sim 10^2\gg 1$. We choose $\omega_\text{max}=3\times 10^{-2}$ so that the low-$k$ condition, $r_+\sqrt{\omega_\text{max}^2-m^2}\sim 10^{-2}\ll 1$, remains satisfied.

Since $1/C(Q_0)=1/Q_0^3$ sets the energy scale below which semiclassical thermodynamics breaks down, smaller values of $C(Q_0)E_i$ correspond to a more extremal BH, for which quantum corrections become increasingly important. This is evident in both Figure~\ref{fig:reflection coefficient} and Figure~\ref{fig:cross section}.

Note that Figure~\ref{fig:reflection coefficient} and Figure~\ref{fig:cross section} exhibit several kinks. In the non-superradiant regime ($\omega>\mu$), these kinks mark the transition where the emission rate per unit frequency $\Gamma_\text{emit}$ drops from nonzero to zero. This feature arises from the $\Theta(E_i-\varpi)$ in the final BH density of states in \eqref{eqn:emission rates}, and therefore appears at $\omega=\mu+E_i$. Similar kinks also occur in the absorption cross-section for neutral scalars (at $\omega=E_i$ instead); see \cite{Emparan:2025sao} for details. 

When $E_i<\mu$, additional kinks appear in the superradiant regime ($m<\omega<\mu$). These mark the point where the absorption rate per unit frequency $\Gamma_\text{abs}$ switches from zero to nonzero due to the $\Theta(E_i+\varpi)$ in the final BH density of states in \eqref{eqn:absorption rates}. Consequently, these kinks occur at $\omega=\mu-E_i$. There is no neutral analog of these kinks\footnote{Note that this is distinct from the ``quantum transparency'' of neutral spinning waves discussed in \cite{Emparan:2025qqf}, where for $\omega<E_{0,\ell}-E_i$ (see subsection~\ref{subsec:Comments on Higher Partial Waves} for the definition of $E_{0,\ell}$ and additional details on spinning waves), the absorption cross-section vanishes identically. In contrast, in our case the absorption rate drops to zero while the (stimulated) emission rate remains nonzero, rendering the absorption cross-section negative.}.

As shown in Figure~\ref{fig:reflection coefficient}, the quantum-corrected reflection coefficient is enhanced relative to its semiclassical counterpart in the superradiant regime, but suppressed outside it. Consequently, the quantum-corrected absorption cross-section is larger than the semiclassical one both inside and outside the superradiant regime, as illustrated in Figure~\ref{fig:cross section}.

Note that as $\omega \to m$, the $\ell=0$ absorption cross-section diverges, which is a characteristic feature of Coulomb-like cross-sections. As already pointed out by Gibbons \cite{Gibbons:1975kk}, in practice, this poses no issue since one always considers a wave packet of incident particles with a finite energy spread. For presentation purposes, we plot $\sigma_\text{abs}(\omega)$ over the interval $[3 \times 10^{-3}, \omega_\text{max}]$ instead of the full range $[m, \omega_\text{max}]$. As we will see, although the absorption cross-section for higher partial waves ($\ell \geq 1$) is significantly suppressed compared to the $\ell = 0$ case, it vanishes smoothly as $\omega \to m$.

\subsection{Higher partial waves}
\label{subsec:Comments on Higher Partial Waves}

Our quantum-corrected s-wave absorption cross-section generalizes straightforwardly to higher partial waves. We will not repeat the same analysis here, but simply provide the results. After absorbing or emitting a higher partial wave component $\phi_{\ell m}$, a BH with initial angular momentum $j_i$ transitions to a final state with angular momentum $j_f$, where $\abs{j_i-\ell}\leq j_f\leq j_i+\ell$. We will focus on the simplest case: $j_i=0$ and $j_f=\ell$.

The total absorption cross-section is given by
\begin{equation}
    \sigma_\text{abs}^\text{tot}=\sum_{\ell=0}^\infty\sigma_\text{abs}^\ell(\omega)\,,
\end{equation}
where each $\sigma_\text{abs}^\ell(\omega)$ denotes the partial wave contribution:
\begin{equation}
    \sigma_\text{abs}^\ell(\omega)=\frac{(2\ell+1)\pi}{\omega^2-m^2}(1-\abs{\mathcal{R_\ell}}^2)\,.
\end{equation}
The quantum-corrected reflection coefficient $\abs{\mathcal{R_\ell}}^2$ is
\begin{equation}
\label{eqn:general ell DOS times matrix element}
\begin{split}
    &1-\abs{\mathcal{R_\ell}}^2=4\mathcal{N}_\ell^2 \abs{\frac{c_{\omega \ell m}}{a_{\omega \ell m}}}^2(2\ell+1)\frac{C}{(2C)^{2\Delta_\ell}\Gamma(2\Delta_\ell)}\\
    &\times\bigg(\sinh(2\pi\sqrt{2C\left(E_i+\varpi -E_{0,\ell}\right)})\Gamma\left(\Delta_\ell\pm i\sqrt{2C(E_i+\varpi-E_{0,\ell})}\pm i\sqrt{2CE_i}\right)\\
    &\hphantom{\times\bigg(} -\sinh(2\pi\sqrt{2C\left(E_i-\varpi -E_{0,\ell}\right)})\Gamma\left(\Delta_\ell\pm i\sqrt{2C(E_i-\varpi-E_{0,\ell})}\pm i\sqrt{2CE_i}\right)\bigg)\,.
\end{split}
\end{equation}
For notational simplicity, we have written $C=C(Q_0)$ and have defined
\begin{equation}
    E_{0,\ell}=\frac{\ell(\ell+1)}{2C(Q_0)},\quad \mathcal{N}_\ell^2 \abs{\frac{c_{\omega \ell m}}{a_{\omega \ell m}}}^2=\frac{2^{4(\Delta_\ell-1)}}{\pi^2}\frac{\Gamma(\Delta_\ell)^2\abs{\Gamma(\Delta_\ell+m_3)}^2}{\Gamma(2\Delta_\ell-1)^2\Gamma(2\Delta_\ell)}(k r_+^2)^{2\Delta_\ell-1}e^{i\pi m_3}\,.
\end{equation}
In \eqref{eqn:general ell DOS times matrix element}, $E_i=M_i-M_0(Q_i)$ is still the initial excitation energy of the BH with charge $Q_i$ and zero angular momentum. After absorbing/emitting the $\phi_{\ell m}$ mode, the final BH energy becomes $M_f=M_i\pm\omega$. However, the extremal energy for a BH with charge $Q_f=Q_i\pm e$ and angular momentum $\ell$ shifts to $M_0(Q_f)+E_{0,\ell}$ \cite{Heydeman:2020hhw}. Therefore, the final excitation energy of the BH is given by
\begin{equation}
    E_f=M_f-M_0(Q_f)-E_{0,\ell}=E_i\pm \varpi -E_{0,\ell}\,.
\end{equation}
Since the BH states are labeled by such excitations \cite{Brown:2024ajk}, this justifies the replacement $E_i\pm \varpi\to E_i\pm \varpi-E_{0,\ell}$ in \eqref{eqn:quantum-corrected absorption cross-section}, leading to \eqref{eqn:general ell DOS times matrix element}. For a complete derivation showing that \eqref{eqn:general ell DOS times matrix element} is indeed the correct generalization for any $\ell$, see \cite{Emparan:2025qqf}.

\begin{figure}[htbp]
    \centering
    \begin{subfigure}[b]{0.49\textwidth}
        \centering
        \includegraphics[width=\textwidth]{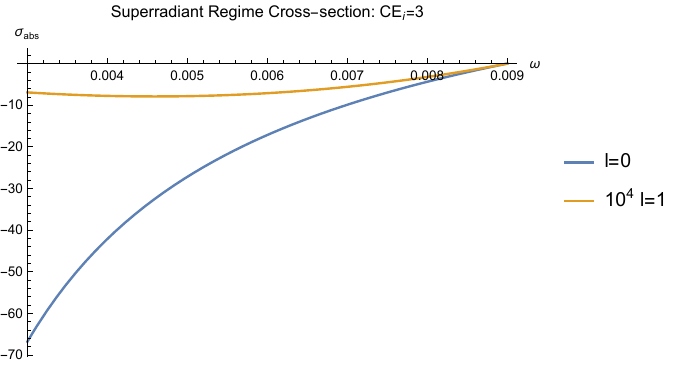}
        \caption{$\sigma^0_\text{abs}$ and $10^4\cdot\sigma^1_\text{abs}$ over the superradiant regime $[3\times10^{-3},\mu)$.}
    \end{subfigure}
    \hfill
    \begin{subfigure}[b]{0.49\textwidth}
        \centering
        \includegraphics[width=\textwidth]{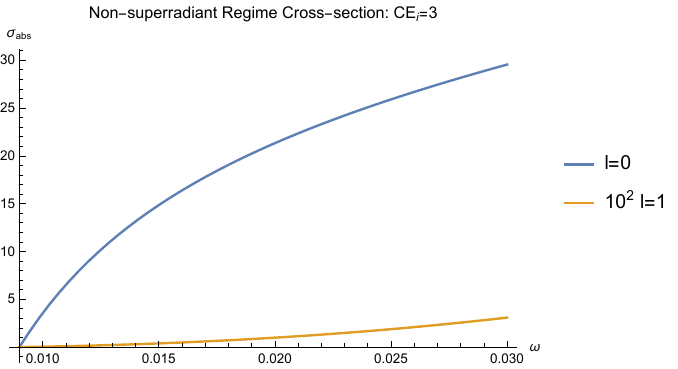}
        \caption{$\sigma^0_\text{abs}$ and $10^2\cdot\sigma^1_\text{abs}$ outside the superradiant regime.}
    \end{subfigure}
    \caption{The quantum-corrected absorption cross-section $\sigma_\text{abs}^\ell$ for $\ell=0$ (blue) and $\ell=1$ (orange) with $C(Q_0)E_i=3$. For clarity in presentation, $\sigma_\text{abs}^1$ is scaled by a factor of $10^4$ in the superradiant regime and by $10^2$ outside it.}
    \label{fig:higher ell cross section}
\end{figure}

As illustrated in Figure~\ref{fig:higher ell cross section}, the higher partial wave contributions are strongly suppressed relative to the s-wave. This provides an \textit{a posteriori} justification for mainly focusing on the s-wave sector throughout this work. 

Nevertheless, a qualitatively different feature of the $\ell \geq 1$ absorption cross-section is that it vanishes as $\omega \to m$. This is reflected in the upward turn of the $\ell=1$ curve as $\omega$ approaches $3 \times 10^{-3} > m$, whereas the $\ell=0$ curve bends downward and diverges to negative infinity as $\omega \to m$.

One could also examine the ``quantum transparency'' property discussed in \cite{Emparan:2025qqf} for the $\ell \geq 1$ absorption cross-section. Because of the $\Theta(E_i + \varpi - E_{0,\ell})$ factor in the final BH density of states after absorption, absorption can occur only if $\omega > E_{0,\ell} + \mu - E_i$, assuming $E_{0,\ell} + \mu - E_i > m$. Similarly, the $\Theta(E_i - \varpi - E_{0,\ell})$ factor implies that emission is possible only if $\omega < E_i + \mu - E_{0,\ell}$. If $E_{0,\ell} > E_i + \mu$, then emission is forbidden and absorption is also prohibited in the interval $[m, E_{0,\ell} + \mu - E_i]$. Consequently, the absorption cross-section vanishes identically in this region. However, given the strong suppression of higher partial waves relative to the s-wave, this effect is not physically significant in our setting.

\section{Discussion}
\label{sec:Discussions}

In this work, we derived the quantum-corrected absorption cross-section \eqref{eqn:quantum-corrected absorption cross-section} for a charged scalar on an (asymptotically flat) RN BH background. As a byproduct, we studied the one-loop correction to the near-extremal RN grand canonical ensemble in detail. Let us first briefly summarize our new findings. 

The one-loop correction to the near-extremal RN-AdS grand canonical ensemble is well understood \cite{Iliesiu:2020qvm}, but no similar analysis has been carried out for the near-extremal RN grand canonical ensemble. We highlighted the need for such a study by showing that a naive extrapolation of the RN-AdS formulae to flat space yields incorrect results. Using the BTZ grand canonical ensemble as an analytic and well understood example, we demonstrated that the (temperature-dependent) one-loop correction to the near-extremal RN grand canonical partition function arises solely from the extremal Schwarzian zero modes. In particular, we resolved a puzzle regarding the expected extremal $U(1)$ gauge zero modes: while these modes have an appreciable contribution in the near-extremal RN-AdS grand canonical ensemble, our analysis in section~\ref{subsec:The Grand Canonical Ensemble} shows—in consistency with the criterion proposed in \cite{Kolanowski:2024zrq}—that their contribution for RN is temperature independent and does not affect the near-extremal behavior to leading order. This insight allowed us to write down the effective 1D action that captures the relevant one-loop correction to the RN grand canonical partition function. We then used the exactly computable matter correlator in this effective 1D theory to determine the quantum-corrected absorption cross-section of the RN BH. Previous studies \cite{Emparan:2025sao, Biggs:2025nzs, Emparan:2025qqf} analyzing similar quantum corrections focused on neutral fields, whereas our analysis enables us to study the scattering and absorption of electrically charged fields both in and out of the superradiant regime. The quantum effects were found to work as amplifiers of the classical results regarding the reflection coefficient (in the superradiant regime) and absorption cross-section (in both superradiant and non-superradiant regimes). They also introduce additional physical features in the form of kinks whereby absorption/stimulated emission is turned off in the supperradiant/non-superradiant regime, respectively.

In this work, we have assumed that the AdS$_2$ BF bound \eqref{eqn:BF bound} is satisfied. It would be interesting to explore what happens when this bound is violated. In that case, the scaling dimension of $\mathcal{O}$ in the AdS$_2$ boundary dual theory becomes complex. A similar observation was made in \cite{Maulik:2025hax}, where the authors interpreted the resulting complex scaling dimension as a novel type of instability, distinct from superradiance. It would be worthwhile to investigate this scenario further. Physically we expect that the system will behave in a similar fashion to a holographic superconductor in holography~\cite{Hartnoll:2008kx,Hartnoll:2008vx}---due to the instability, the charged scalar will condense in the near-horizon region forming a charged cloud outside the BH (a ``hairy BH'') or the final state could resemble a charged ``star'' with no horizon. It would also be interesting to understand the role that quantum effects have in this instability in the context of the AdS/CFT correspondence, in order to clarify better the physics of quantum critical points at small temperature and finite chemical potential\footnote{In a sense this means zooming in and understanding better the physics of quantum effects near the so-called ``dome region'', see~\cite{keimer2015quantum} for a concensus on the actual phase diagram for copper oxides and~\cite{Zaanen:2021llz} for a theoretical review of such quantum critical points.}.

Another interesting, though more formal, direction would be to gain a better understanding of the temperature-independent one-loop corrections arising from extremal gauge or rotational zero modes. In the BTZ case, we have explicitly shown that these modes do not affect the leading near-extremal one-loop corrected grand canonical partition function, which is fully captured by the extremal Schwarzian zero modes alone. For the RN case, however, we were not able to carry out an explicit calculation; instead, we extrapolated the BTZ result and neglected the temperature-independent one-loop contribution from the extremal gauge zero modes. It would be reassuring if future work could confirm that neglecting their contribution is indeed justified. We hope to report further results in these directions soon.

As a final note, it is well known that in four dimensional asymptotically flat space notorious IR divergences pose difficulties in properly defining an S-matrix. It would be very interesting to explore the physical interplay between the near-extremal low energy quantum corrections, with the IR divergences of the S-matrix due to the flat asymptotics and the long range nature of the electromagnetic and gravitational forces.

\acknowledgments

We wish to thank Elias Kiritsis, Thomas Mertens, and Gustavo Joaquin Turiaci for discussions. P.B. acknowledges financial support from the European Research Council (grant BHHQG-101040024), funded by the European Union. Views and opinions expressed are those of the authors only and do not necessarily reflect those of the European Union or the European Research Council. Neither the European Union nor the granting authority can be held responsible for them.

\appendix


\section{The density of states and matter correlators from the Schwarzian}
\label{subsec:The Density of States and Matter Correlators}

For convenience, we collect in this appendix the quantum-corrected BH density of states and the transition amplitudes and rates between different BH states, as computed using the 1D effective Schwarzian theory that governs the dynamics of the soft quantum fluctuations near extremality (zero temperature).

The density of states as a function of energy and charge defined by
\begin{equation}
    \mathcal{Z}_\text{RN}[\beta,\mu]=\sum_{Q\in e\cdot\mathbb{Z}}e^{\beta\mu \frac{Q}{e}}\int_{M_0(Q)}^\infty\diff M\,e^{-\beta M}\rho(M,Q)\,,
\end{equation}
can be obtained from \eqref{eqn:RN grand partition function from partition function} by an inverse Laplace transform to be
\begin{equation}
    \rho(E,Q)=\Theta(E)\frac{e^{S_0(Q)}C(Q)}{2\pi^2}\sinh(2\pi\sqrt{2C(Q)E})\,,
\end{equation}
where $\Theta$ is the Heaviside step function and we have defined $E=M-M_0(Q)$ to be the excitation above the extremal energy of a charged-$Q$ BH with energy $M$.

Consider the emission of a charged-$e$ particle with energy $\omega$. The changes in the BH's energy and charge are
\begin{equation}
    M_f-M_i=-\omega,\quad Q_f-Q_i=-e\,.
\end{equation}
As a result, the change in the BH's excitation level is\footnote{If we were considering an RN-AdS BH, then using \eqref{eqn:corrected RN-AdS grand canonical energy due to n}, we would find
\begin{equation}
\label{eqn:difference between initial and final RN-AdS BH excitations}
    E_f-E_i=-\left(\omega-\mu-\frac{q_i+q_f}{2K_\text{RN-AdS}}\right)\,,
\end{equation}
where $e\cdot q=Q-\Tilde{Q}_0$. Thus, \eqref{eqn:difference between initial and final RN BH excitations} can be understood as the $K_\text{RN-AdS}\to\infty$ limit of \eqref{eqn:difference between initial and final RN-AdS BH excitations}, consistent with the fact that $K_\text{RN}$ diverges.}
\begin{equation}
\label{eqn:difference between initial and final RN BH excitations}
    E_f-E_i=-\left(\omega-\frac{e}{\sqrt{4\pi G_N}}\right)=-(\omega-\mu)\,.
\end{equation}

Let $\mathcal{O}$ be an operator with scaling dimension $\Delta$ and charge $e$ in the 1D effective theory \eqref{eqn:1D effective theory for RN grand canonical}. The square of its matrix element between BH states $\ket{E_i,Q_i}$ and $\ket{E_f,Q_f}$ is 
\begin{equation}
    \abs{\braket{E_f,Q_f|\mathcal{O}|E_i,Q_i}}^2=\delta_{Q_i,Q_f+e}\cdot 2e^{-S_0(Q_0)}\frac{\Gamma\left(\Delta\pm i\sqrt{2C(Q_0)E_f}\pm i\sqrt{2C(Q_0)E_i}\right)}{(2C(Q_0))^{2\Delta}\Gamma(2\Delta)}\,,
\end{equation}
where the Kronecker delta is the matrix element for the $U(1)$ holonomy $e^{i\alpha}$ \cite{Iliesiu:2019lfc, Brown:2024ajk, Mertens:2018fds}
\begin{equation}
    \delta_{Q_i,Q_f+e}=\int_0^{2\pi}\frac{\diff \alpha}{2\pi}\,(D^{Q_i}(e^{i\alpha}))^*D^{e}(e^{i\alpha})D^{Q_f}(e^{i\alpha}),\quad D^{Q}(e^{i\alpha})=e^{iQ\alpha}\,.
\end{equation}
The remaining factor is the standard Schwarzian matrix element \cite{Mertens:2017mtv, Yang:2018gdb, Iliesiu:2019xuh, Mertens:2022irh}. We have adopted the useful notation that whenever $\pm$ appears inside a gamma function, it means one should take a product of that gamma function with both signs.

Finally if we assume that background BH charge is much larger than the charge fluctuations: $\abs{Q_f-Q_i}\ll Q_i\sim Q_f\sim Q_0$. Then 
\begin{equation}
\label{eqn:density of states times matrix element}
    \begin{split}
        \rho(E_f,Q_f)\abs{\braket{E_f,Q_f|\mathcal{O}|E_i,Q_i}}^2\approx& \delta_{Q_i,Q_f+e}\frac{C(Q_0)}{\pi^2}\sinh(2\pi\sqrt{2C(Q_0)E_f})\\
        &\times \frac{\Gamma\left(\Delta\pm i\sqrt{2C(Q_0)E_f}\pm i\sqrt{2C(Q_0)E_i}\right)}{(2C(Q_0))^{2\Delta}\Gamma(2\Delta)}\,,
    \end{split}
\end{equation}
\eqref{eqn:density of states times matrix element} will be used to compute the transition rate between the initial and final BH states.

\newpage
\bibliographystyle{JHEP}
\bibliography{bib}

\end{document}